\documentclass[12pt,preprint]{aastex}
\usepackage{lscape}

\begin{document}

\newcommand{\OI}{~1}
\newcommand{\NaIa}{~2}
\newcommand{\NaIb}{~3}
\newcommand{\MgI}{~4}
\newcommand{\AlI}{~5}
\newcommand{\SiIGa}{~6}
\newcommand{\SiISn}{~4}
\newcommand{\CaI}{~7}
\newcommand{\CaISn}{~4}
\newcommand{\ScII}{12}
\newcommand{\TiIa}{~8}
\newcommand{\TiIb}{~9}
\newcommand{\TiIc}{10}
\newcommand{\TiISn}{~4}
\newcommand{\TiII}{11}
\newcommand{\MnI}{12}
\newcommand{\FeIa}{13}
\newcommand{\FeIb}{14}
\newcommand{\FeIc}{15}
\newcommand{\FeIOB}{16}
\newcommand{\FeIBKa}{17}
\newcommand{\FeIBKb}{18}
\newcommand{\FeII}{19}
\newcommand{\NiI}{20}
\newcommand{\NiISn}{~4}
\newcommand{\CuI}{21}
\newcommand{\BaII}{~5}
\newcommand{\LaII}{~4}
\newcommand{\EuII}{22}

\title{CHEMICAL ABUNDANCES FOR SEVEN GIANT STARS IN M68 (NGC~4590) :
A GLOBULAR CLUSTER WITH ABNORMAL SILICON AND TITANIUM ABUNDANCES}

\author{Jae-Woo Lee\altaffilmark{1}, Bruce W.\ Carney\altaffilmark{2,3}
\and Michael J.\ Habgood\altaffilmark{2}}

\altaffiltext{1}{Department of Astronomy \& Space Science,
Astrophysical Research Center for the Structure and
Evolution of the Cosmos,
Sejong University, 98 Gunja-Dong, Gwangjin-Gu, Seoul, 143-747;
jaewoo@arcsec.sejong.ac.kr}
\altaffiltext{2}{Department of Physics \& Astronomy,
University of North Carolina, Chapel Hill, NC 27599-3255;
bruce@physics.unc.edu,mjames@astro.unc.edu}
\altaffiltext{3}{Visiting Astronomer, Cerro Tololo Inter-American Observatory,
National Optical Astronomy Observatories,
which are operated by the Association of
Universities for Research in Astronomy, Inc., under contract with the
National Science Foundation.}

\begin{abstract}

We present a detailed chemical abundance study of seven giant stars in M68
including six red giants and one post-AGB star. We find significant
differences in the gravities determined using photometry and those
obtained from ionization balance, which suggests that non-LTE
effects are important for these low-gravity, metal-poor stars.
We adopt an iron abundance using photometric gravities and
Fe~II lines to minimize those effects, finding
[Fe/H] = $-$2.16 $\pm$ 0.02 ($\sigma = 0.04$). 
For element-to-iron ratios,
we rely on neutral lines vs.\ Fe~I and ionized lines vs. Fe~II
(except for [O/Fe]) to also minimize non-LTE effects. 

We find
variations in the abundances of sodium among the program stars.
However, there is no correlation (or anti-correlation) with the
oxygen abundances. Further, the post-AGB star has a normal (low)
abundance of sodium. Both of these facts add further support to
the idea that the variations seen among some light elements
within individual globular clusters arises from primordial
variations, and not from deep mixing.

M68, like M15, shows elevated abundances of silicon compared
to other globular clusters and comparable metallicity field
stars. But M68 deviates even more in showing a relative
underabundance of titanium. We speculate that in M68,
titanium is behaving like an iron-peak element rather than
its more commonly observed adherence to enhancements seen
in the ``$\alpha$" elements such as magnesium, silicon,
and calcium. We interpret this result as implying that
the chemical enrichment seen in M68 may have arisen from
contributions from supernovae with somewhat more massive
progenitors than contribute normally to abundances seen
in other globular clusters.

The neutron capture elements barium and europium vary
among the stars in M15 (Sneden et al.\ 1997), but
the [Ba/Eu] is relatively constant, suggesting that
both elements arise in the same nucleosynthesis events.
M68 shares the same [Ba/Eu] ratio as the stars in M15,
but the average abundance ratio of these elements,
and lanthanum, are lower in M68 relative to iron than
in M15, implying a slightly weaker contribution of 
$r$-process nucleosynthesis in M68. 

\end{abstract}


%==========================================================================
\section{INTRODUCTION}
\begin{figure}
\epsscale{1}
\figurenum{1}
\plotone{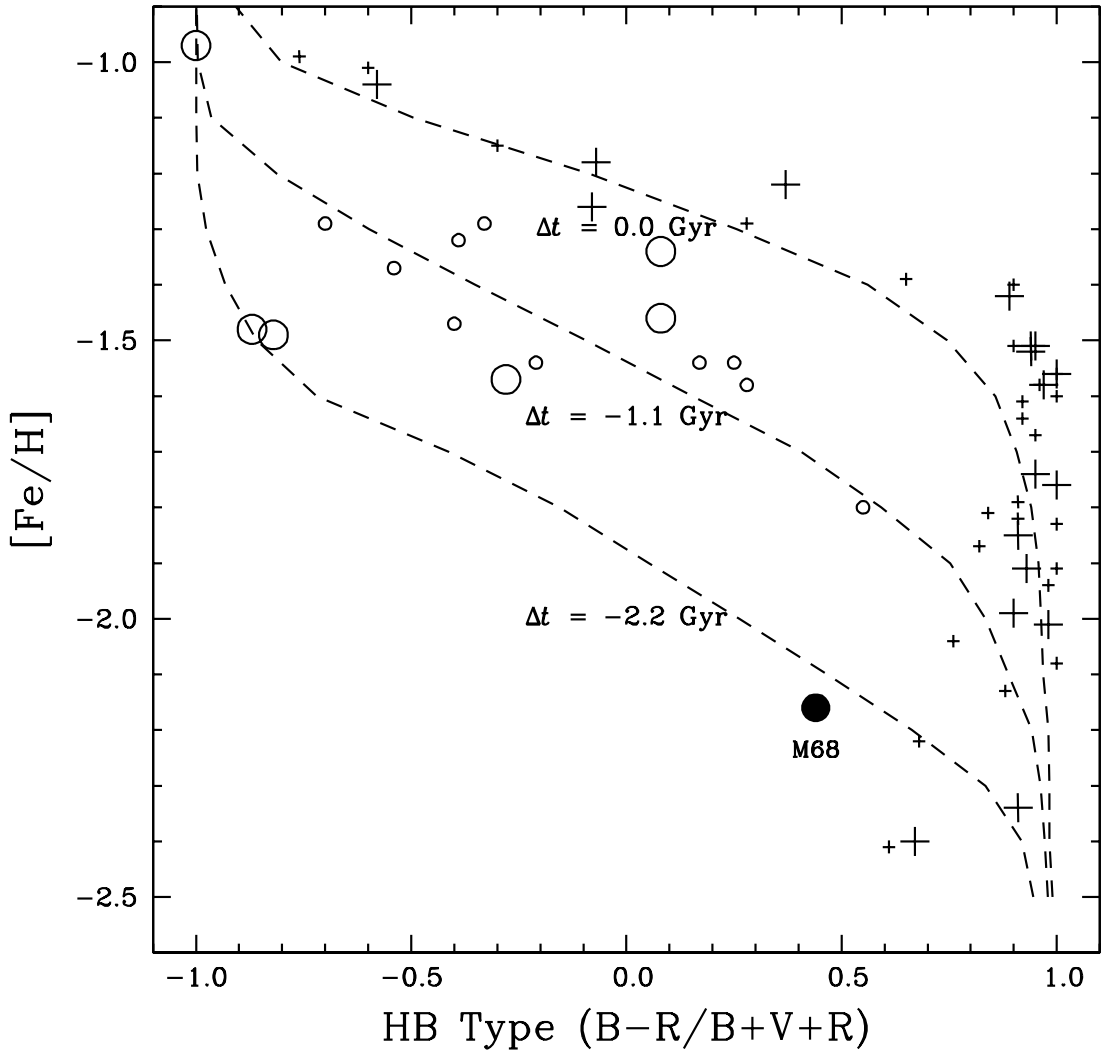}
\caption{HB type versus [Fe/H] (HB isochrones) of ``halo" globular clusters
(Da Costa \& Armandroff 1995) and M68.
Crosses are ``old halo" clusters and open circles are ``younger halo" clusters,
where large crosses and open circles denote clusters studied employing
high-resolution spectroscopy (see also Figure~2).
HB isochrones for $\Delta t$ = 0.0, $-$1.1, and $-$2.2 Gyr
(with respect to the mean age of ``old halo" globular clusters)
are also shown with dashed lines (Rey et al.\ 2001).\label{fig:hbiso}}
\end{figure}

Globular clusters in our Galaxy do not define a single homogeneous
population with a single history. In his pioneering work,
Zinn (1985) subdivided the globular clusters into the ``halo"
and the ``thick disk" groups at [Fe/H] = $-$0.8.
The halo clusters have an essentially spherical distribution about
the Galactic center and they constitute a pressure supported system
(a small rotational velocity and a larger velocity dispersion),
while the thick disk clusters have a highly flattened spatial distribution
and constitute a rotational supported system
(a larger rotational velocity and a smaller velocity dispersion).
Searle \& Zinn (1978) and Lee, Demarque, \& Zinn (1994)
suggested that the inner halo globular  clusters exhibit
a tight horizontal branch (HB) morphology versus [Fe/H] relation,
while the outer halo globular clusters show the second parameter
phenomenon (i.e., a larger scatter in HB
type\footnote{HB type is defined to be (B$-$R)/(B+V+R), where
B, V, R are the numbers of blue HB stars, RR Lyrae variables,
red HB stars, respectively (Lee, Demarque, \& Zinn 1990).} at a given [Fe/H]).
Subsequently, Zinn (1993) subdivided the halo clusters into two groups.
The ``old halo" group obeys the same HB type versus [Fe/H] relationship
as the inner halo clusters while the ``younger halo" group deviates from
this relationship by a significant amount (see Figure~\ref{fig:hbiso}).
Zinn (1993) and Da Costa \& Armandroff (1995) argued that the old and
the younger halo groups have different kinematic properties that
the old halo group has a prograde mean rotation velocity with
a smaller velocity dispersion while the younger halo group has a retrograde
mean rotation velocity about the Galactic center with a larger velocity
dispersion. They suggested that the old halo group formed during the collapse
that led ultimately to the formation of the Galactic disk and
the younger halo group were accreted later in time.
If so, one would expect to see signatures of different chemical enrichment
history carved in spectra of stars in globular clusters.
For example, one might be able to investigate 
the early form of the initial mass function (IMF) by
studying abundances of stellar mass-sensitive elements
(see discussion and references in McWilliam 1997).

An alternative perspective on ``younger halo" and ``old halo"
clusters is that of dissolved and accreted dwarf galaxies.
For example, Freeman (1993) suggested that the massive and
chemically unusual globular cluster $\omega$~Cen might be
the remnant nucleus of an accreted dwarf galaxy. Lynden-Bell \&
Lynden-Bell (1995) noted the possible alignments of orbital
poles of some globular clusters such that they might comprise
a ``spoor" (more commonly referred to as a stream) of clusters.
While the specific details have not found support, the idea
has been demonstrated very nicely by Dinescu et al.\ (2000),
who found a clear dynamical relationship between the
space motions of the young globular cluster Palomar~12
and the Sagittarius dwarf galaxy. Indeed, other globular clusters
may also be associated with this particular accretion event
(NGC 5634: Bellazzini, Ferraro, \& Ibata 2002;
NGC~4147: Bellazzini et al.\ 2003; Palomar~2: Majewski et al.\ 2004).
Yoon \& Lee (2002) have similarly speculated on a dynamical
relationship between some metal-poor globular clusters,
including M68. Yoon \& Lee (2002) suggested that M68 was a member
of a satellite galaxy and accreted later in time to our Galaxy,
resulting in a planar motion of several metal-poor clusters
including M15 and M92.

Dynamical evolution in the Galaxy can eventually dissolve
streams, making their detection difficult. However, an interesting
alternative exists, called ``chemical tagging". As Freeman \&
Bland-Hawthorn (2002) discussed, stars born in galaxies whose
star formation histories differ from those that have created the
bulk of the Galaxy's stars may still be discernible in
unusual element-to-iron ratios. Indeed, Cohen (2004) has
found a compelling link between Palomar~12 and the Sagittarius
dwarf. Detailed chemical abundances of globular clusters
may yet become the principle means of identifying historical
links between stars and clusters that are now widely dispersed
in our Galaxy.

M68 (NGC~4590) is part of the younger halo, according to its HB type
with respect to its metallicity, and is located 10 kpc from the Galactic center.
Dinescu, Girard, \& van~Altena (1999) have found that the cluster's
Galactic orbit carries it rather far from the Galactic center,
perhaps as far as 30 kpc.
As shown in Figure~\ref{fig:hbiso}, globular clusters with [Fe/H] $\leq$ $-$2.0
have HB types that are greater than 0.6 with the exception of M68.
If age is the second parameter\footnote{The ``second parameter"
is the disputed variable that, in addition to mean metallicity,
determines the HB type of a cluster.}, M68 represents the low metallicity
tail of the younger halo globular clusters.
The absolute age of M68 does appears to be slightly younger than
those of the oldest globular clusters in our Galaxy.
Rosenberg et al.\ (1999) argued that M68 is coeval to or
slightly younger than those of the oldest halo globular clusters while
VandenBerg (2000) discussed that the age of M68 is less than
that of M92, one of the oldest globular clusters in our Galaxy,
by $\approx$ 15\%.

The previous metallicity estimates of M68 as follows.
Zinn \& West (1984) and Zinn (1985) adopted
[Fe/H] = $-$2.09.
Gratton \& Ortolani (1989) observed two stars in M68 with
one being in common with this study. They derived elemental
abundances for 11 species including iron, finding [Fe/H] = $-$1.92.
However, due to the low resolving power ($R$ = 15,000) of their spectra,
their equivalent width measurements were vulnerable to line blending.
Minniti et al.\ (1993) observed two stars in M68,
one being in common with Gratton \& Ortolani (1989) and this study,
finding [Fe/H] = $-$2.17.
Minniti et al.\ (1996) also discussed oxygen and 
sodium abundances of the cluster.
Finally, Rutledge (1997) observed the infrared Ca~II triplet for 19 stars
and obtained [Fe/H] = $-$2.11 $\pm$ 0.03 using the 
Zinn \& West (1984) [Fe/H] scale.

In this paper, we explore the detailed elemental abundances for
seven giant stars in M68.
One of these stars is a probable asymptotic giant
branch (AGB) stars and the remaining six are red giant branch (RGB) stars.
This study is tied directly to that of Lee \& Carney (2002) with
the same instrument setups and analysis methods.

%==========================================================================
\section{OBSERVATIONS AND DATA REDUCTION}

\begin{figure}
\epsscale{1}
\figurenum{2}
\plotone{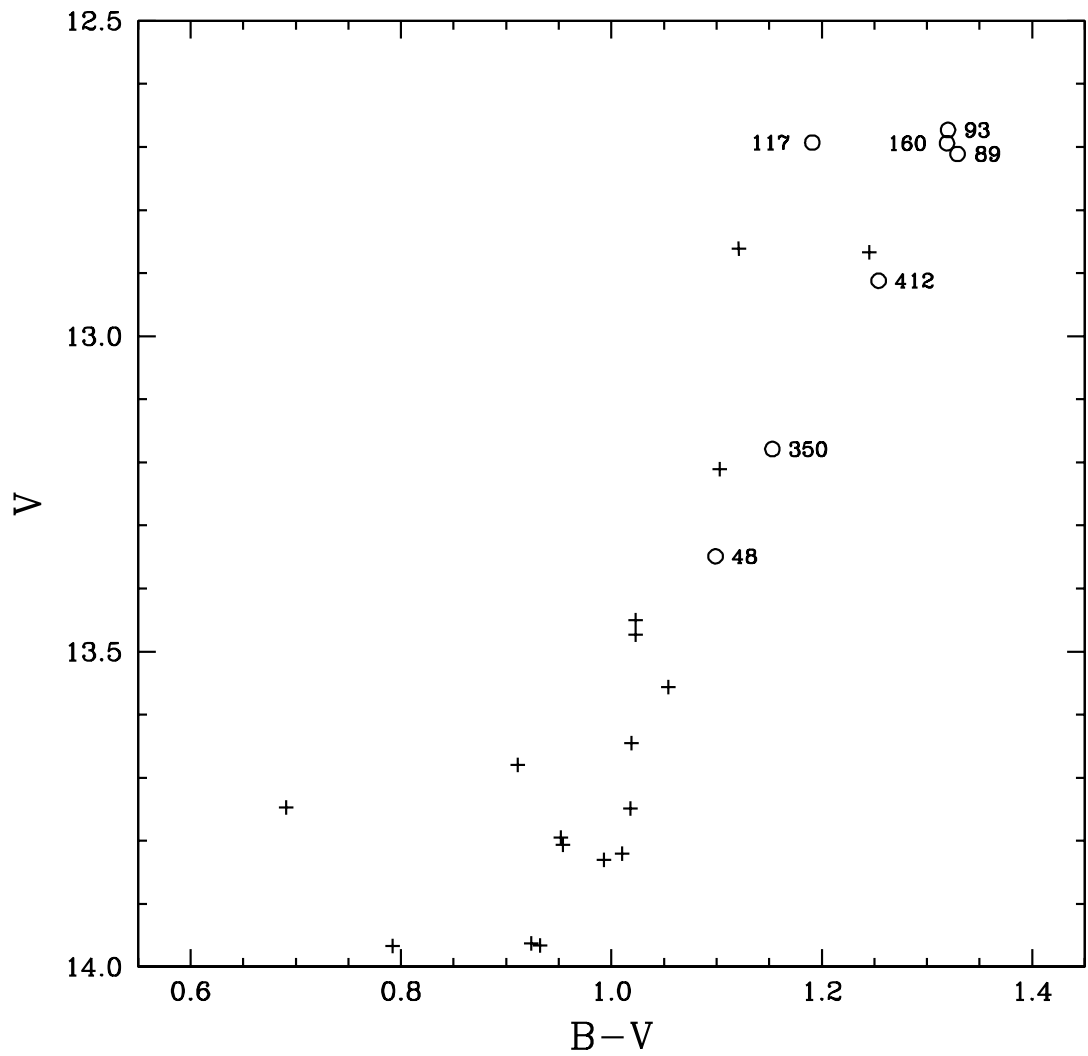}
\caption{Color-magnitude diagram of M68 with photometry
from Walker (1994), showing the positions of our program stars.\label{fig:cmd}}
\end{figure}

The observations were carried out from 4 to 7 May 1996.
We selected our program RGB stars from the $BV$ photometry of Walker (1994).
The positions of our target giant stars on the color-magnitude diagram
along with bright stars in M68 are shown in Figure~\ref{fig:cmd}.
In Table~\ref{tab:obs}, we provide identifications
(Alcaino 1977; Harris 1975; Walker 1994),
position (Cutri et al.\ 2000),
$V$ magnitudes, $(B-V)$ colors (Walker 1994) and $K$ magnitudes
(Frogel, Persson, \& Cohen 1983; Cutri et al.\ 2000) of our target stars.
Please note that the 2MASS $K$ magnitudes 
have been converted to the CIT system
(Cutri et al.\ 2000) are in good agreement with those of Frogel et al.\ (1983).
We obtained high signal-to-noise ratio (S/N $\geq$ 90 per pixel) echelle spectra
using the CTIO 4-meter telescope and its Cassegrain echelle spectrograph.
The Tek 2048 $\times$ 2048 CCD, 31.6~lines/mm echelle grating,
long red camera, and G181 cross-disperser were employed
for our observations.
The slit width was 150 $\mu$m, or about 1.0 arcsec, that projected to
2.0 pixels and which yielded an effective resolving power $R$ = 28,000.
Each spectrum had complete spectral coverage from 5420 to 7840~\AA.
All program star observations were accompanied by flat lamp,
Th-Ar lamp, and bias frames.

The raw data frames were trimmed, bias-corrected, and flat-fielded
using the IRAF\footnote{IRAF (Image Reduction and Analysis
Facility) is distributed by the National Optical Astronomy
Observatory, which is operated by the Association of Universities
for Research in Astronomy, Inc., under contract with the National
Science Foundation.} ARED and CCDRED packages. The scattered light
was also subtracted using the APSCATTER task in ECHELLE package.
The echelle apertures were then extracted to form 1-d spectra, which
were continuum-fitted and normalized, and a wavelength solution
was applied following the standard IRAF echelle reduction
routines.

Equivalent widths were measured mainly by the direct integration of
each line profile using the SPLOT task in IRAF ECHELLE package.
We estimate our measurement error in equivalent width to be
$\pm$2 m\AA\ from the size of  noise features in the spectra and our ability
to determine the proper continuum level.
The equivalent widths for our program stars are listed in Table~\ref{tab:ew}.

\begin{figure}
\epsscale{1}
\figurenum{3}
\plotone{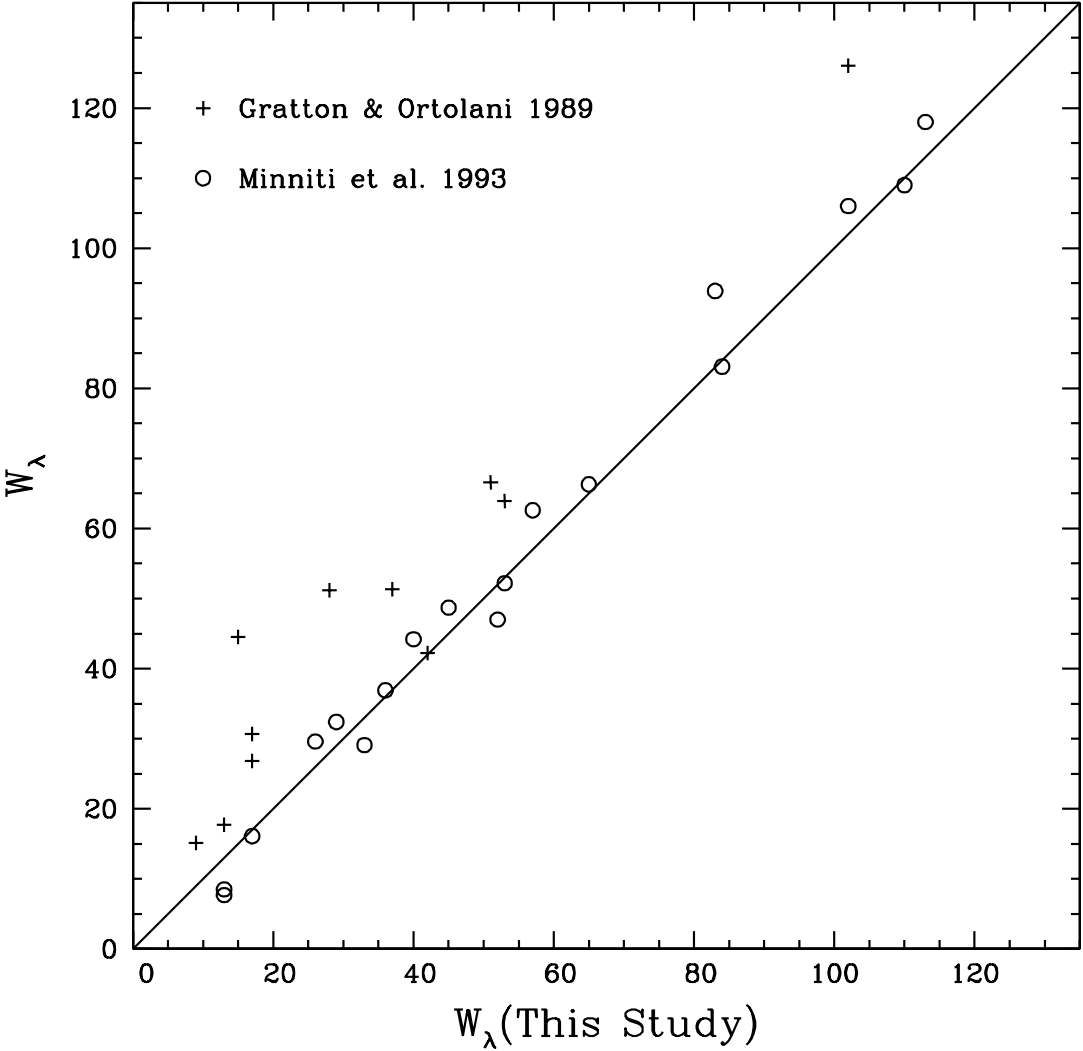}
\caption{A comparison of equivalent widths of the star 160
with those of Gratton \& Ortolani (1989) and Minniti et al\ (1993).\label{fig:compew}}
\end{figure}

Gratton \& Ortolani (1989) and Minniti et al.\ (1993) obtained
a spectrum of one star in common with our program stars.
They used the identifications from Harris (1975)
while we have used those of Walker (1994).
Comparing the two shows that the star I-260 from Harris (1975)
is the same as the star 160 from Walker (1994).
Figure~\ref{fig:compew} compares equivalent width measurements of our
work with those measured by Gratton \& Ortolani (1989;
crosses) and Minniti et al.\ (1993; open circles).
The agreement with the latter study is quite good,
with a mean difference of $2.3 \pm 1.5$ m\AA\ (in the
sense of their study minus ours). The instrumental
resolving powers for the two studies were very
similar (27,000 vs.\ 28,000). However, the lower
resolving power of the observations reported by
Gratton \& Ortolani (1989), about 15,000, appears to
have led to systematically larger equivalent widths,
as Figure~\ref{fig:compew} reveals.
The mean difference is $13.8 \pm 2.7$ m\AA.

%======================================================================
\section{ANALYSIS}
In our elemental abundance analysis, we use the usual spectroscopic
notations that
$[A/B] \equiv \log (N_A/N_B)_{star} - \log (N_A/N_B)_{\sun}$,
and that $\log n(A) \equiv \log (N_A/N_H) + 12.00$ for each element.
For the absolute solar iron abundance, we adopt $\log n$(Fe) = 7.52
following the discussion of Sneden et al.\ (1991).

\subsection{Line Selection and Oscillator Strengths}

For our line selection, laboratory oscillator strengths
were adopted whenever possible,
with supplemental solar oscillator strength values.
In addition to oscillator strengths, taking into account
the damping broadening due to the van der Waals force, we adopted
the Uns\"old approximation with no enhancement.

The abundance analysis depends mainly on the reliability
of the oscillator strength values of the Fe~I and Fe~II lines,
since not only the metallicity scale but also the
stellar parameters, including the spectroscopic temperature, surface
gravity, and microturbulent velocity, will be determined using
these lines. As discussed by Lee \& Carney (2002),
we mainly relied upon the extensive laboratory
oscillator strength measurements by the Oxford group
(Blackwell et al.\  1982b, 1982c, 1986a).
We also used oscillator strength values measured by O'Brian et al.\ (1991)
and the Hannover group (Bard, Kock, \& Kock 1991; Bard \& Kock 1994).
In our iron abundance analysis, we consider the Oxford group's measurements
(the absorption method) as the ``primary" oscillator strengths and
oscillator strength measurements that relied on emission methods
(O'Brian et al.\ 1991; Bard, Kock, \& Kock 1991; Bard \& Kock 1994)
as ``secondary". Therefore, the oscillator strengths by O'Brian et al.\
and the Hannover group were scaled with respect to those by
the Oxford group as a function of excitation potential
(de Almeida 2000, private communication),
\begin{eqnarray}
\log gf &=& \log gf(OB) - 0.017, \nonumber \\
\log gf &=& \log gf(H91) - 0.015 - 0.009 \chi, \nonumber \\
\log gf &=& \log gf(H94) - 0.027 - 0.009 \chi,
\label{eqn6:gfscale}
\end{eqnarray}
where the excitation potential $\chi$ is given in electron volts.
Blackwell, Smith, \& Lynas-Gray (1995) also pointed out that there
appears to exist a slight gradient in the excitation potential
between oscillator strengths by the Oxford group and those by the
Hannover group, with $\log gf(Oxford) = \log gf(Hannover) - 0.021
- 0.006\chi$.

For neutral titanium lines, we relied on the laboratory
measurements by the Oxford group (Blackwell et al.\ 1982a, 1983,
1986b). It should be noted that the original Oxford $gf$-values
have been increased by +0.056~dex following Grevesse, Blackwell,
\& Petford (1989). They discussed that the Oxford $gf$-values
relied on the inaccurate lifetime measurements and the absolute
$gf$-values should be revised based on the new measurements.

Hyperfine splitting (HFS) components must be considered
in the barium abundance analysis because Ba~II lines are
usually very strong even in metal-poor stars and
the desaturation effects due to HFS components become evident
(see for example, McWilliam 1998).
We adopted the Ba~II HFS components and
oscillator strengths of Sneden et al.\ (1997).
We also perform HFS treatment for scandium, manganese 
(Prochaska \& McWilliam 2000) and copper (Kurucz 1993).
For the copper HFS analysis,
we adopt the solar Cu isotopic ratio, 69\% $^{63}$Cu and 31\% $^{65}$Cu,
following the discussions given by Smith et al.\ (2000) and
Simmerer et al.\ (2003).
The equivalent widths of Mn~I $\lambda$ 6021.79 \AA\ and
Cu~I $\lambda$ 5782.13 \AA\ lines
in our program stars are weak and the Mn and Cu abundance differences between
HFS treatment and non-HFS treatment are no larger than 0.02 dex.
We list our source of oscillator strengths for each element
in Table~\ref{tab:gf}.

\subsection{Stellar Parameters and Model Atmospheres}

Having good stellar parameters, such as the effective temperature and
the surface gravity, is critical for any stellar abundance study,
since the absolute or the relative elemental abundance scale will depend
on the input stellar parameters.
For our analysis, we rely on spectroscopic temperatures and
photometric surface gravities, following the method described
in Lee \& Carney (2002; see also Ivans et al.\ 2001,
Kraft \& Ivans 2003, and Sneden et al.\ 2004).
It has been suspected by others that the traditional spectroscopic surface
gravity determination method which requires the same elemental abundances
derived from neutral and singly ionized lines
(preferentially Fe~I and Fe~II lines)
suffers from non-local thermodynamic equilibrium (NLTE) effects
(see, for example, Nissen et al.\ 1997; Allende Prieto et al.\ 1999).
Since metal-poor stars have much weaker metal-absorption in the ultraviolet,
more non-local UV flux can penetrate from the deeper layers.
This flux is vital in determining the ionization equilibrium of the atoms,
resulting in deviations from local thermodynamic equilibrium (LTE).
Nissen et al.\ (1997) claimed that surface gravities of metal-poor
dwarfs and subgiants derived from the spectroscopic method, which demands
that Fe~I and Fe~II lines should provide the same iron abundance,
are a factor of two or three ($\Delta \log g \approx$ 0.3 -- 0.5)
smaller than those from the Hipparcos parallaxes.
Allende Prieto et al.\ (1999) also claimed that spectroscopic gravities and
those from the Hipparcos parallaxes are in good agreement for stars
in the metallicity range $-1.0 <$ [Fe/H] $<$ +0.3, while
large discrepancies can be found for stars with metallicities below
[Fe/H] = $-$1.0, in the sense that the spectroscopic method provides
lower surface gravities. Therefore, we rely on photometric gravities
for our abundance analysis.

The initial estimates of the temperature of program stars were estimated
using $BVK$ photometry of our program stars
(Cutri et al.\ 2000; Frogel, Persson, \& Cohen 1983; Walker 1994)
and the empirical color-temperature relations given by
Alonso, Arribas, \& Martinez-Roger (1999).
Since their relation depends slightly on the metallicity,
we adopted [Fe/H] = $-$2.1 for M68 (Harris 1996).
To estimate the dereddened color, we adopt $E(B-V)$ = 0.07 (Walker 1994).
Note that Alonso et al.\ (1999) employed the Carlos S\'anchez Telescope (TCS)
system for their infrared color-temperature relations
and we used the relation given by Alonso, Arribas, \& Martinez-Roger (1998)
to convert the CIT system to the TCS system  for $K$ magnitudes
listed in Table~\ref{tab:obs}.
To derive photometric surface gravity in relation to that of the Sun,
we use $\log g_{\sun}$ = 4.44 in cgs units,
$M_{\rm bol,\sun}$ = 4.74 mag, and $T_{\rm eff,\sun}$ = 5777 K for the Sun
(Livingston 1999) and we assume the stellar masses for
all the red giants stars in this analysis to be $M$ = 0.8 $M_\sun$.
We use the empirical relation given by Alonso et al.\ (1999)
to estimate the bolometric correction and we adopt $(m-M)_0$ = 14.97 mag
for the cluster (Harris 1996).

With initial photometric temperature and surface gravity estimates,
72-depth plane-parallel LTE model atmospheres were computed
using the program ATLAS9, written and supplied by Dr.\ R.\ L.\ Kurucz.
Assuming that all of the cluster's stars would prove to be metal-poor,
the model atmospheres were computed using opacity distribution functions
and abundances with enhanced abundances of all the ``$\alpha$" elements
(O, Ne, Mg, Si, S, Ar, Ca, and Ti) by 0.4~dex.
The ``$\alpha$" element enhancements are important since
several of these elements are quite abundant and are major
electron donors to the H$^{-}$ opacity.
During our model computation, convective overshoot was turned off.

The abundance analysis was performed using the current version (2002)
of the LTE line analysis program MOOG (Sneden 1973).
Adopting the photometric temperature and surface gravity
as our initial values, we began by restricting the analysis to
those Fe~I lines with $\log$(W$_{\lambda}$/$\lambda$) $\leq$ $-5.2$
(i.e., for  the linear part of the curve of growth),
and comparing the abundances as a function of excitation potential.
New model atmospheres were computed with a slightly different effective
temperature until the slope of the log~n(Fe~I)  versus excitation potential
relation was zero to within the uncertainties.
The stronger Fe~I lines were then added and the microturbulent
velocity $v_{\rm turb}$ altered until the $\log$ n(Fe~I) versus
$\log$(W$_{\lambda}$/$\lambda$) relation had no discernible slope.

Table~\ref{tab:pam} shows our temperature and surface gravity of program stars.
In the fifth column of the Table, we also show the temperature
of five stars in common given by Frogel et al.\ (1983).
Please note that Gratton \& Ortolani relied on temperatures given
by Frogel et al.\ (1983), which are 159 $\pm$ 27 K
higher than our spectroscopic temperatures.
The temperature difference between those derived from $(B-V)$ colors
(column 2) and spectroscopic temperatures of this study is
72 $\pm$ 15 K (7 stars), in the sense that our spectroscopic
temperature is low.
The temperature difference between those derived from $(V-K)$ colors
using $K$ magnitudes of Frogel et al.\ (column 3) and
spectroscopic temperatures of this study is
54 $\pm$ 17 K (5 stars) and
those derived from $(V-K)$ colors
using 2MASS $K$ magnitudes (column 4) and
spectroscopic temperatures of this study is
69 $\pm$ 14 K (7 stars).
The discrepancy between photometric and spectroscopic temperatures
from $(V-K)$ colors is slightly smaller than that from $(B-V)$ colors.

%======================================================================
\section{RESULTS}

\subsection{Elemental Abundances and Error Analysis}

\begin{figure}
\epsscale{1}
\figurenum{4}
\plotone{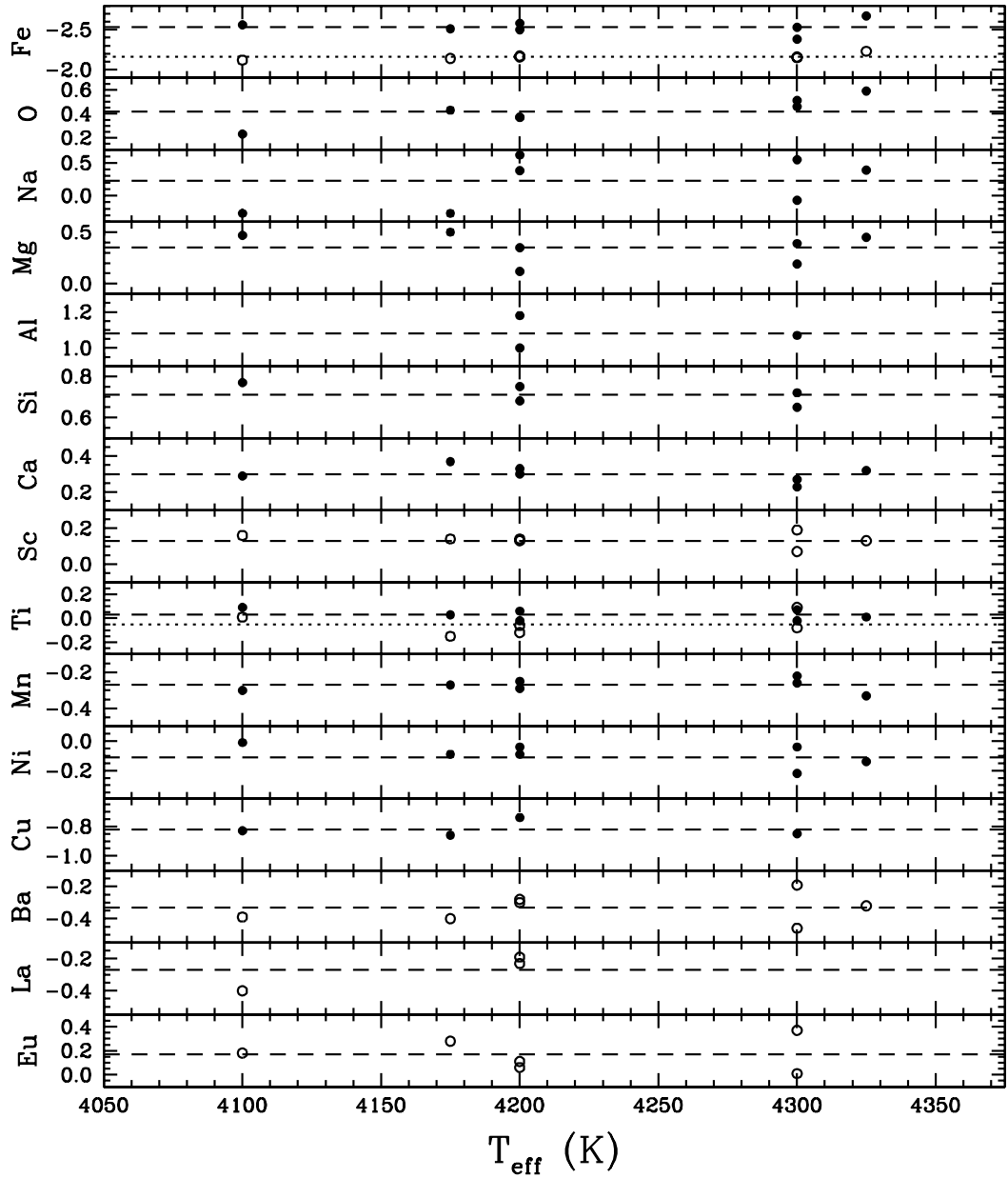}
\caption{Elemental abundances studied in this work plotted
against $T_{\rm eff}$. Note that the ordinate of the top panel
is [Fe/H], while the ordinates of the remaining panels
are [el/Fe] ratios. Neutral atoms are presented by filled circles
and singly ionized atoms by open circles.
Dashed lines are the mean abundances of each element.
The mean Fe~II and Ti~II abundances are presented by dotted lines.\label{fig:abundteff}}
\end{figure}

In Tables~\ref{tab:abundph}, we present the elemental abundances of
our program stars using photometric surface gravities
and spectroscopic temperatures.
The [el/Fe] ratios for neutral elements are estimated from [el/H]
and [Fe~I/H] ratios, with the exception of oxygen.
The [el/Fe] for singly ionized elements (Ti~II, Ba~II, La~II, and Eu~II)
and oxygen are estimated from [el/H] and [Fe~II/H] ratios
(see, for example, Ivans et al.\ 2001, Kraft \& Ivans 2003,
Sneden et al.\ 2004).
The internal uncertainty quoted is for a single line and, therefore,
that of each element is given by $\sigma/\sqrt{n}$,
where $\sigma$ is the uncertainty per line and
$n$ is the number of absorption lines used for each element.
Systematic errors, such as in adopted $gf$ values as a function
of excitation potential, which could lead to systematically
erroneous temperature estimates, are not included.
The last two columns of the Table show the mean values of each element
of the cluster with and without the star 117 (ZNG2),
an ultraviolet-bright, post-AGB star 
according to Zinn, Newell, \& Gibson (1972).
We adopt iron abundances based only on the Fe~II lines
for our program stars since the Fe~II abundance
is thought to be less sensitive to NLTE conditions
(see for example, Th\'evenin \& Idiart 1999, Kraft \& Ivans 2003).
In Figure~\ref{fig:abundteff}, we show abundances of all elements
measured in this study against $T_{\rm eff}$, showing no discernible
gradient in elemental abundances with  $T_{\rm eff}$.
The mean [Fe/H] of $-$2.16 dex for our seven stars is measured
with a small internal uncertainty of $\pm$0.02 ($\sigma$ = 0.04).

For comparison, we also show the elemental abundances of our
program stars using the traditional spectroscopic surface
gravities (see column 7 of Table~\ref{tab:pam}).
In the Table, the [el/Fe] ratios are derived from
[el/H] and mean [Fe/H] ratios.
In Table~\ref{tab:diff}, we show differences in
elemental abundances using photometric and spectroscopic gravities.
The [Fe/H] ratio using spectroscopic gravities is 0.25 dex
lower than that using photometric gravities.
For other elements, however, the elemental abundances from
two different methods are in good agreement to within 0.10 dex.
Therefore, these small differences suggest that
we are still able, in principle, to compare our elemental abundances 
with other results
using the different surface gravity determination method. Nonetheless,
we reiterate that we choose to employ photometric gravities
and will discuss only the results from Table~\ref{tab:abundph}.

In Table~\ref{tab:moddep}, we show estimated errors resulting from
uncertainties in the input model atmosphere $\delta T_{\rm eff}$ = $\pm$ 80 K,
$\delta \log g$ = $\pm$0.3 and $\delta v_{\rm turb}$ = $\pm$0.2 km s$^{-1}$,
which are appropriate for our analysis.
The Fe, Si, Ba, and La abundances are sensitive to $T_{\rm eff}$ resulting in
$\mid$$\delta$[el/Fe (or H for Fe)]/$\delta T_{\rm eff}$(80 K)$\mid$
$\approx$ 0.08 -- 0.12 dex.
The iron abundance from Fe~II lines is sensitive to surface gravity,
$\mid$$\delta$[Fe~II/H]/$\delta \log g$(0.3 dex)$\mid$ $\approx$ 0.10 dex.
Since our program stars are metal-poor, absorption lines are usually weak
as listed in Table~\ref{tab:ew}.
Therefore, our derived elemental abundances are less sensitive to
the microturbulent velocity. The barium abundance is the most
sensitive to the microturbulent velocity resulting in
$\mid$$\delta$[Ba/Fe]/$\delta v_{\rm turb}$(0.2 km s$^{-1}$)$\mid$
$\approx$ 0.09 dex.

\subsection{Comparisons with previous results}

As mentioned above, Gratton \& Ortolani (1989) and
Minniti et al.\ (1993, 1996) derived abundances for
star 160 (I-260 from Harris 1975).
In Table~\ref{tab:comp160}, we compare our stellar parameters and
elemental abundances of this star with those of
Gratton \& Ortolani (1989) and Minniti et al. (1993, 1996).

Gratton \& Ortolani (1989) relied on the results from Frogel et al.\ (1983)
for the estimated temperature and surface gravity.
Table~\ref{tab:pam} shows that this results in
a $T_{\rm eff}$ estimate 230~K hotter than our spectroscopic measurement,
and a slightly lower microturbulent velocity (by 0.2 km s$^{-1}$) as well.
Our results suggest that a difference in microturbulent velocity of
0.2 km s$^{-1}$  does not alter the derived iron abundances
and element-to-iron ratios by more than about 0.1 dex.
As discussed above (see also Figure~\ref{fig:compew}),
the equivalent widths measured by  Gratton \& Ortolani
are about 14 m\AA\ larger than our measurements.
This was also noticed by Minniti et al.\ (1993),
who found a difference in equivalent width of 10.2 $\pm$ 9.0 m\AA,
in the sense of Gratton \& Ortolani minus Minniti et al.
Therefore, the larger equivalent widths with the warmer surface temperature
are the probable cause of the higher [Fe/H] value they derived.
Therefore, the detailed comparison of each elemental abundance
between this study and Gratton \& Ortolani may not be meaningful.

Minniti et al.\ (1993) derived an even higher temperature of the star,
$T_{\rm eff}$ =  4400~K, which is 300~K warmer than our temperature, and,
consequently, their iron abundance of the star from Fe~I lines is
about 0.45 dex higher than our value, but, interestingly, very
close to the value we obtain using only the Fe~II lines.
Minniti et al.\ (1993) obtained $\log g$ = 1.0
for the star, using the ionization equilibrium of iron lines,
which is about 0.3 dex larger than  those of Gratton \& Ortolani (1989)
and this study.
This is very worrisome, since Minniti et al.\ (1993) used
the same temperature determination method, the same model atmospheres
as this study and similar oscillator strength values.
Their instrumental resolution was comparable to ours, and they
employed the program WIDTH rather than MOOG, but both programs
yield the same results, in our experience, at least
when the atomic data are identical. The equivalent widths
are also in good agreement, as noted earlier.
In particular, the agreement in the equivalent widths of weak lines
(i.e., the lines on the linear part of the curve of growth)
appears to be excellent.
For lines with $\log$(W$_{\lambda}$/$\lambda$) $\leq$ $-5.2$,
the difference in equivalent widths is $-$1.7 $\pm$ 1.4 m\AA\ (7 lines),
in the sense of Minniti et al.\ minus this study.

\begin{figure}
\epsscale{1}
\figurenum{5}
\plotone{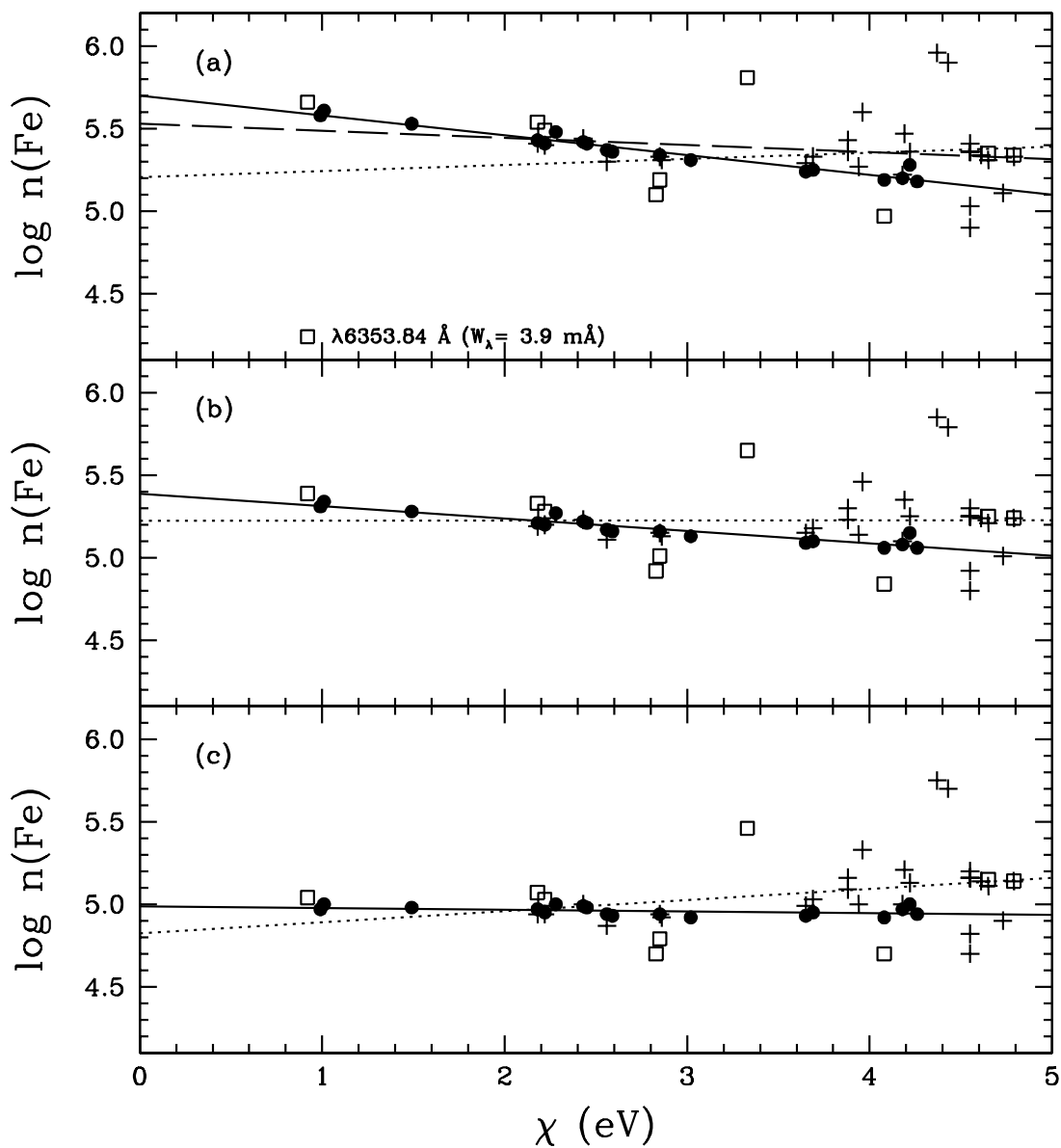}
\caption{(a) The $\log$~n(Fe) versus excitation potential relation
of the star M68-160 (I260) using Fe~I equivalent widths
of Minniti et al.\ (1993) with
$\log$(W$_{\lambda}$/$\lambda$) $\leq$ $-5.2$  for
$T_{\rm eff}$ = 4400 K,  $\log g$ = 1.0, and $v_{\rm turb}$ = 2.0 km s$^{-1}$.
Crosses are for Fe~I lines with W$_{\lambda}$ $>$ 10 m\AA\ and
open squares are for W$_{\lambda}$ $\leq$ 10 m\AA\ from Minniti et al.\ (1993).
The dotted and the dashed lines represent the least square fits to the
data with and without the Fe~I $\lambda$ 6538.84 \AA, respectively.
Filled circles show the relation using Fe~I equivalent widths of
this study and the solid line is for the least square fit to the data.
(b) Same as (a), but using the model atmosphere with
$T_{\rm eff}$ = 4275 K,  $\log g$ = 1.0, and $v_{\rm turb}$ = 2.0 km s$^{-1}$
and the Fe~I $\lambda$ 6538.84 \AA\ line is not used.
(c) Same as (b), but using the model atmosphere with
$T_{\rm eff}$ = 4100 K,  $\log g$ = 0.7, and $v_{\rm turb}$ = 1.8 km s$^{-1}$.
\label{fig:compminniti}}
\end{figure}

Instead of comparing our results to those of Minniti et al.\ (1993) directly,
we have rederived the surface temperature using
the equivalent widths measured by Minniti et al.\ (1993).
In their Table~2\footnote{The titles of column 2 and 3
in Table~2 of Minniti et al.\ (1993) should be switched.},
Minniti et al.\ (1993) presented Fe~I and Fe~II line widths of the star 160.
Using their Fe~I line widths with $\log$(W$_{\lambda}$/$\lambda$) $\leq$ $-5.2$,
we ran MOOG for this star again. At this time, we use a Kurucz
model atmosphere with $T_{\rm eff}$ = 4400 K, $\log g$ = 1.0,
and $v_{\rm turb}$ = 2.0 km s$^{-1}$ as our initial input model.
Note that these stellar parameters are derived by Minniti et al.\
for their analyses of the star as shown in Table~\ref{tab:comp160}.
We show the results from our MOOG run in Figure~\ref{fig:compminniti}a.
We obtain [Fe/H] = $-$2.18 $\pm$ 0.29 (36 lines) and this value
is very similar to that of Minniti et al., [Fe/H] = $-$2.11.
As can be seen in the Figure, however, the scatter in our analysis
using the line list of Minniti et al.\ is much larger than
that Minniti et al.\ estimated ($\sigma$ $\leq$ 0.07 dex per line)
and we were not able to reproduce their results using their input data.
We find it necessary to adopt a different temperature.
Since the Fe~I line at $\lambda$ 6353.84\AA\ deviates
far from the mean value of the relation, we excluded this line
and ran MOOG again with $T_{\rm eff}$ = 4275 K and obtain
[Fe/H] = $-$2.30 $\pm$ 0.22 dex (Figure~\ref{fig:compminniti}b).
For comparison, we show the results from MOOG run with
our stellar parameters, $T_{\rm eff}$ = 4100 K, $\log g$ = 0.7,
and $v_{\rm turb}$ = 1.8 km s$^{-1}$
in Figure~\ref{fig:compminniti}c. We obtain [Fe/H] = $-$2.50 $\pm$ 0.13.
In the Figure, the agreement between results from our equivalent widths
and those of Minniti et al.\ for lines with $\chi$ $\leq$ 3 eV is good.
For lines with $\chi$ $>$ 3 eV, the scatter in results from Minniti et al.\
is rather large.
It is likely that the Fe~I lines employed by 
Minniti et al.\ may have been contaminated,
and as a result they would have obtained a higher iron abundance
than we have.
Their accidental inclusion of Fe~I lines suffering from line blending
made their measurements to meet the ionization equilibrium
at high surface gravity $\log~g$ = 1.0.
Using our line width measurements listed in Table~\ref{tab:ew},
the criterion of the ionization equilibrium of iron lines
is not satisfied until the surface gravity becomes as low as
$\log~g$ = 0.0 and this surface gravity value is not only 1 dex smaller
than that of Minniti et al., but is implausible.

We conclude that we do not understand why we cannot obtain the same results
as did Minniti et al.\ (1993) using their data, and we conclude that
the agreement between their [Fe/H] result and ours is probably fortuitous.

\section{DISCUSSION}

In the previous section, we discussed the rationale by which
we have derived elemental abundance
estimates. From the ionized iron lines, we obtain
[Fe/H] = $-$2.16 $\pm$ 0.02 ($\sigma = 0.04$) for M68 (based on internal
errors only). Element-to-iron ratios are matched using comparable
ionization states (except for oxygen, given its very high
ionization potential). 

We explore the abundances of a variety of elements relative to iron.
In the interests of economy and of interest, we focus on comparisons
between M68 and the comparably metal-poor globular cluster M15, based
on the results of Sneden et al.\ (1997, 2000), and then
discuss these in turn compared to other ensembles of field stars
and clusters. We select M15 because
it is one of the few metal-poor globular clusters with comparable data for
the neutron capture elements lanthanum, barium, and europium, and partly
because of the hypothesized common origin (Yoon \& Lee 2002).

\subsection{Mixing or Primordial Variations? O, Na, Mg, Al}

\begin{figure}
\epsscale{1}
\figurenum{6}
\plotone{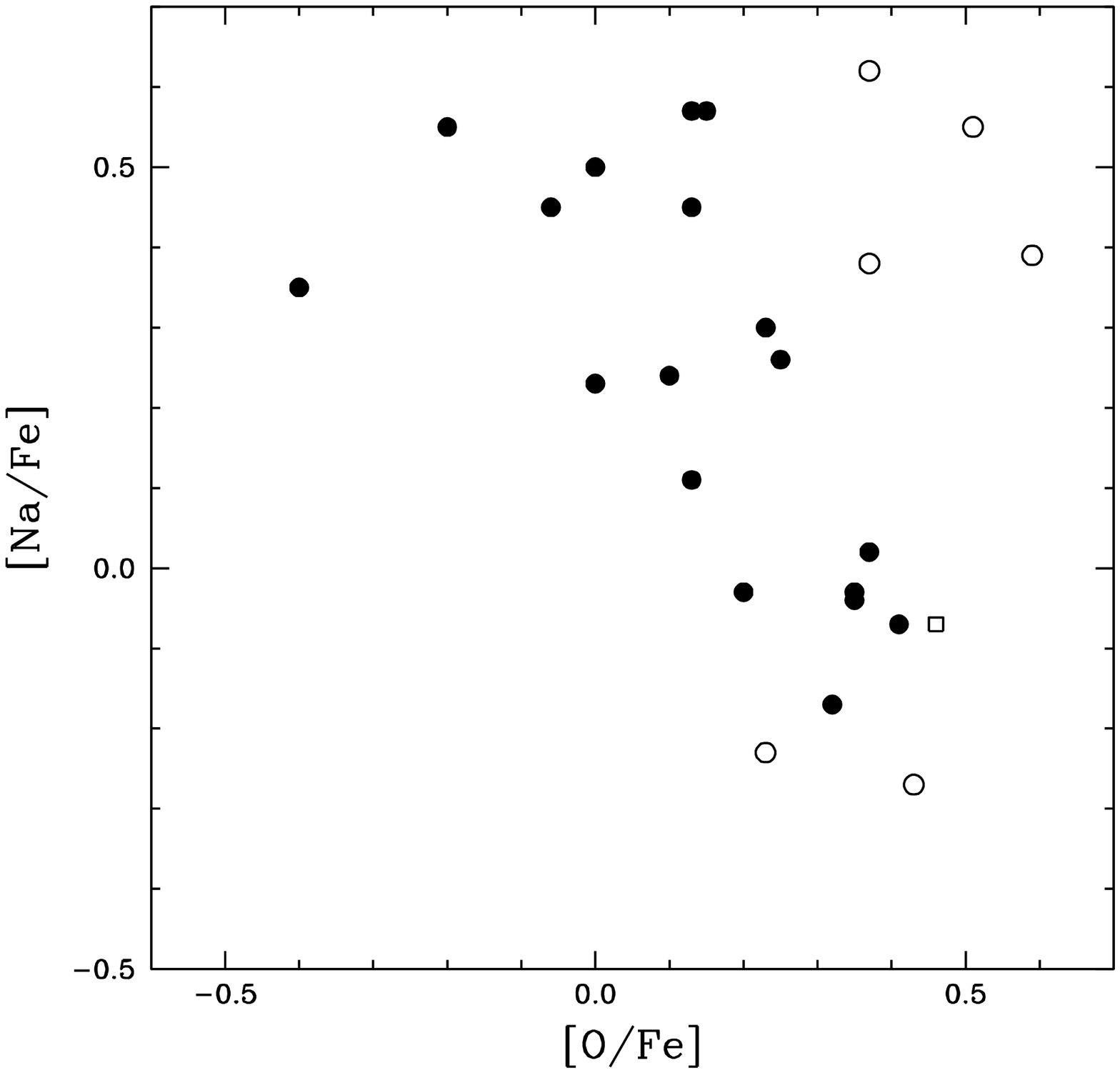}
\caption{A comparison of [Na/Fe] vs.\ [O/Fe] for M15 (dots) vs.\
those of M68 (circles). The post-AGB star M68-117 is
marked with an open square. \label{fig:ona}}
\end{figure}

Many globular clusters appear to show anticorrelations between the
abundances of oxygen and sodium, and of magnesium and aluminum. The
subject was reviewed by Kraft (1994), and has been revisited by numerous
authors. The approach often taken has been that these anticorrelations
arise from deep mixing, whereby material whose chemical compositions have
been altered by proton captures within the CNO cycle have been brought
to the stellar surface. This concept has become less plausible with the
discovery that such anticorrelations are seen in relatively
unevolved stars in the metal-poor clusters NGC~6397 and NGC~6752 
(Gratton et al.\ 2001) and in the metal-rich cluster 47~Tuc (Carretta
et al.\ 2004). The variations seen in some of these elements in
Table~5 nonetheless warrant a quick look at these possible anti-correlations
in M68, and we show them here in Figure~\ref{fig:ona}, in
comparison with the results for M15 from Sneden et al.\ (1997).

There are three relatively interesting results here. First is that
M68 does show a variation, especially in sodium. The second is
that the large range in sodium abundances is \underline{not}
matched by a correlated (or anti-correlated) variation in the
oxygen abundances, as appears to be the case for M15. Finally,
the post-AGB star 117, indicated by an open square in the Figure,
is consistent with no signs of deep mixing, despite it clearly
being the most evolved star in our study. The lack of
an anti-correlation and the lack of variations seen in star 117
are consistent with a primordial variation for the differences
in elemental abundances, presumably due to pollution during the
earliest stages of the cluster formation and evolution by AGB 
nucleosynthesis and mass loss.

\subsection{The Other Light Elements: Si, Ca, and Ti Abundances}

Table~5 shows that the other light elements, silicon, calcium,
and titanium, do not show any detectable variation in their
abundances. These elements are of considerable interest nonetheless
because they provide us with
an opportunity to compare the nucleosynthesis histories of clusters,
and, in principle, a means to compare the rate of star formation
and, possibly, relative ages.

If star formation began everywhere in the Galaxy at about the
same time, then the abundances of elements that emerge from
Type II supernovae, including these light elements and $r$-process
elements such as europium, will be enhanced relative to the
abundances of elements that emerge from nucleosynthesis
sites that appear more gradually, including $s$-process
elements from AGB stars, and iron-peak elements from Type~Ia
supernovae. The details are complex, but this basic picture
is consistent with the basic behavior of these ``$\alpha$" elements'
abundances (see the discussions of Wyse \& Gilmore 1988;
Wheeler et al.\ 1989, and Carney 1996). If we compute 
unweighted averages for silicon, calcium, and titanium,
we find for M68 that
[$\alpha$/Fe] = $+0.34 \pm 0.22$, which is roughly consistent
with that found in M15 and other globular clusters.
In Table~\ref{tab:alpha}, we summarize the [Si/Fe], [Ca/Fe], [Ti/Fe], and
the mean [$\alpha$/Fe] ratios of M68, other globular clusters
(Gratton 1987; Gratton \& Ortolani 1989;
Kraft et al.\ 1992, 1995, 1997, 1998; McWilliam et al.\ 1992;
Sneden et al.\ 1994, 1997, 2000b; Brown et al.\ 1997, 1999;
Ivans et al.\ 1999; Shetrone \& Keane 2000)
and field stars with $-2.50 \leq$\ [Fe/H] $\leq\ -1.90$ 
studied by Fulbright (2000).
(The errors are those of the mean.)
In the Table, old inner halo clusters denote metal-poor
inner halo clusters NGC~6287 ([Fe/H] = $-$2.01),
NGC~6293 ($-$1.99), and NGC~6541 ($-$1.76)
and their Galactocentric distances are
1.6 kpc, 1.4 kpc, and 2.2 kpc, respectively  (Lee \& Carney 2002).
The mean values of [$\alpha$/Fe] do not appear to vary much,
and are probably consistent within the internal errors (the
simple averages) and any lingering systematic errors.

A closer examination of the abundances of these elements does
reveal some interesting differences, however.
The silicon abundance of M68 appears to be similar to that
found in M15, and both are
enhanced relative to those of other clusters and field stars.

The calcium abundance of M68 is likewise similar to those found
in M15, and both are similar to other halo stars and clusters.
There may, however, be a modest difference
($\approx$ 0.15 dex) between M68, M15, and other
``old halo" clusters and ``younger halo" and thick disk clusters.
Note that this contradicts the assignment of M68 to the
``younger halo" category.

\begin{figure}
\epsscale{1}
\figurenum{7}
\plotone{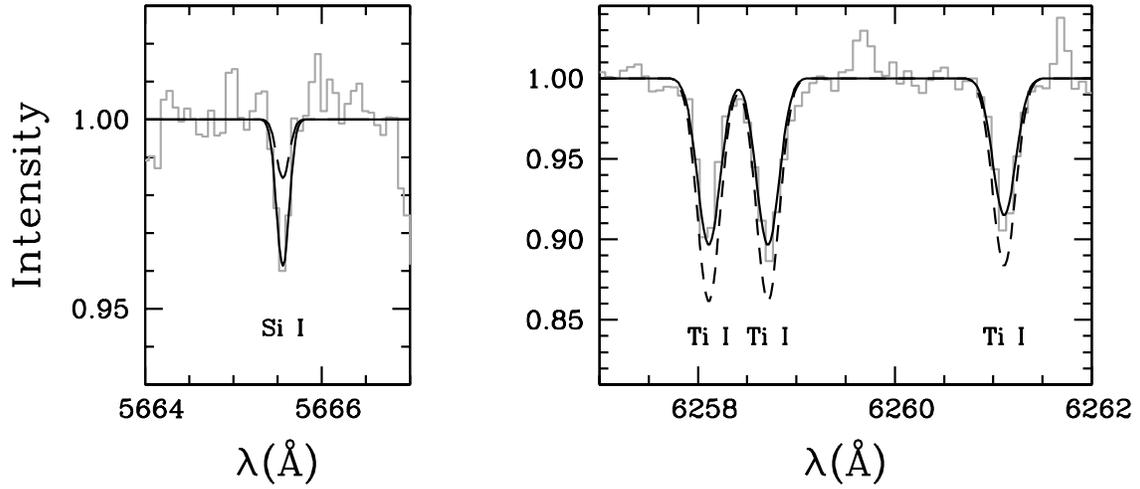}
\caption{Comparisons of observed spectrum of the star M68-93
with those of synthetic spectra near Si~I $\lambda$ 5665.56\AA\
and Ti~I $\lambda$ 6258.11, 6258.71, and 6261.11\AA.
The observed spectrum is presented by histograms,
synthetic spectra with [Si/Fe] = +0.68, [Ti/Fe] = +0.06 by solid lines
(see also Table~5), and synthetic spectra with [Si/Fe] = +0.30,
[Ti/Fe] = +0.30 by dotted lines.\label{fig:synspec}}
\end{figure}

On the other hand, the titanium abundance of M68
appears to be much lower than all other clusters.
Is this effect real?
The titanium abundance of the metal-poor RGB stars
using the neutral titanium lines may suffer from NLTE effects,
such as an over-ionization, and the resultant Ti abundance may be
spurious. However, our Ti abundance analyses using the Ti~II lines
also yield a lower titanium abundance for our program stars,
indicating that they are truly titanium deficient.
In Figure~\ref{fig:synspec}, we show comparisons of
observed spectrum of the star 93 with those of synthetic spectra
near Si~I $\lambda$ 5665.56\AA\
and Ti~I $\lambda$ 6258.11, 6258.71, and 6261.11\AA.
In the Figure, the observed spectrum is presented by histograms,
synthetic spectra with [Si/Fe] = +0.68, [Ti/Fe] = +0.06 by solid lines
(see also Table~\ref{tab:abundph}), and synthetic spectra with [Si/Fe] = +0.30,
[Ti/Fe] = +0.30, which elemental abundances have long been thought
to be {\em normal} (if there exists normal elemental abundances
among globular clusters) for globular clusters, by dotted lines.
The observed spectrum clearly shows that the Si~I absorption line
is too strong to be [Si/Fe] = +0.30, while the Ti~I absorption
lines are too weak to be [Ti/Fe] = +0.30. So the deficiency
appears to be real. 

While we have included this element
because its abundances often track those of the lighter
elements, including oxygen, magnesium, silicon, and calcium,
titanium may also be considered to be an iron-leak
element. Explosive nucleosynthesis calculations of the massive stars
(Woosley \& Weaver 1995) predict that one of the major sources of
the SNe~II titanium yield is $^{48}$Cr via
the consecutive electron capture processes. Further, their
models predicted that
SNe~II with masses in the range 25 - 40 $M_\odot$ are likely
to overproduce Si compared with Ti.
We turn, therefore, to a more detailed comparison between M68
and M15, concentrating on the light element silicon, thought
to be mostly produced in Type~II supernovae, and nickel, mostly
associated with Type~Ia supernovae. 
Sneden et al.\ (1997) also found a high silicon abundance in M15.
Restricting the sample to only those stars with well-determined
abundances, they found
$<$[Si/Fe]$>$ = +0.62 $\pm$ 0.06. Using again only the stars with
well-determined titanium abundances, and comparing [Ti~I/Fe~I]
and [Ti~II/Fe~II], the four stars studied by Sneden et al.\ (1997, 2000)
resulted in $<$[Ti/Fe]$>$ = $+0.27 \pm 0.08$ (the errors are all
given here as errors of the mean). Thus for M15, [Si/Ti] is +0.35,
whereas for M68 it is much higher, +0.76. For the iron-peak element
nickel, our seven stars reveal $<$[Ni/Fe]$>$ = $-0.11 \pm 0.03$ 
($\sigma = 0.08$) dex. We are uncertain how to compare our results
to those from Sneden et al.\ (1997, 2000), however. Sneden et al.\ (1997)
found $<$[Ni/Fe]$>$ = $+0.14 \pm 0.03$, based on 12 stars, which is very different
than what we have found for M68. But Sneden et al.\ (2000) found
$<$[Ni/Fe]$>$ = $-0.21 \pm 0.02$ ($\sigma = 0.04$) dex, based on
three of the same stars studied earlier. Sneden et al.\ (2000)
commented on the difficulties in establishing truly reliable
stellar parameters and abundances, and, alas, this difference
is yet another aspect of that problem.

We conclude our discussion of these elements with the possibility
that the stars we have studied in M68 sampled a relatively high
end of the initial mass function and the resultant supernovae.

\subsection{The Heavy Neutron Capture Elements Ba, La, and Eu}

The elements heavier than the iron-peak elements can not be
efficiently produced by the charged-particle interactions
due to the large Coulomb repulsion between the nuclei,
and they are thought to be produced through both slow ($s$-) and
rapid ($r$-) neutron capture processes.
The $s$-process occurs mainly in low- (1-3 $M_\odot$) or
intermediate-mass (4-7 $M_\odot$) AGB stars,
while the $r$-process is thought to occur in type II supernovae explosions.
Therefore, comparisons between the $r$-process (europium) and
$s$-process elements (barium and lanthanum) provide a clue
to the history of the Galactic nucleosynthesis, since $r$- and
$s$-processes are thought to occur in stars with very different
masses and, therefore, in different evolutionary timescales.

\begin{figure}
\epsscale{1}
\figurenum{8}
\plotone{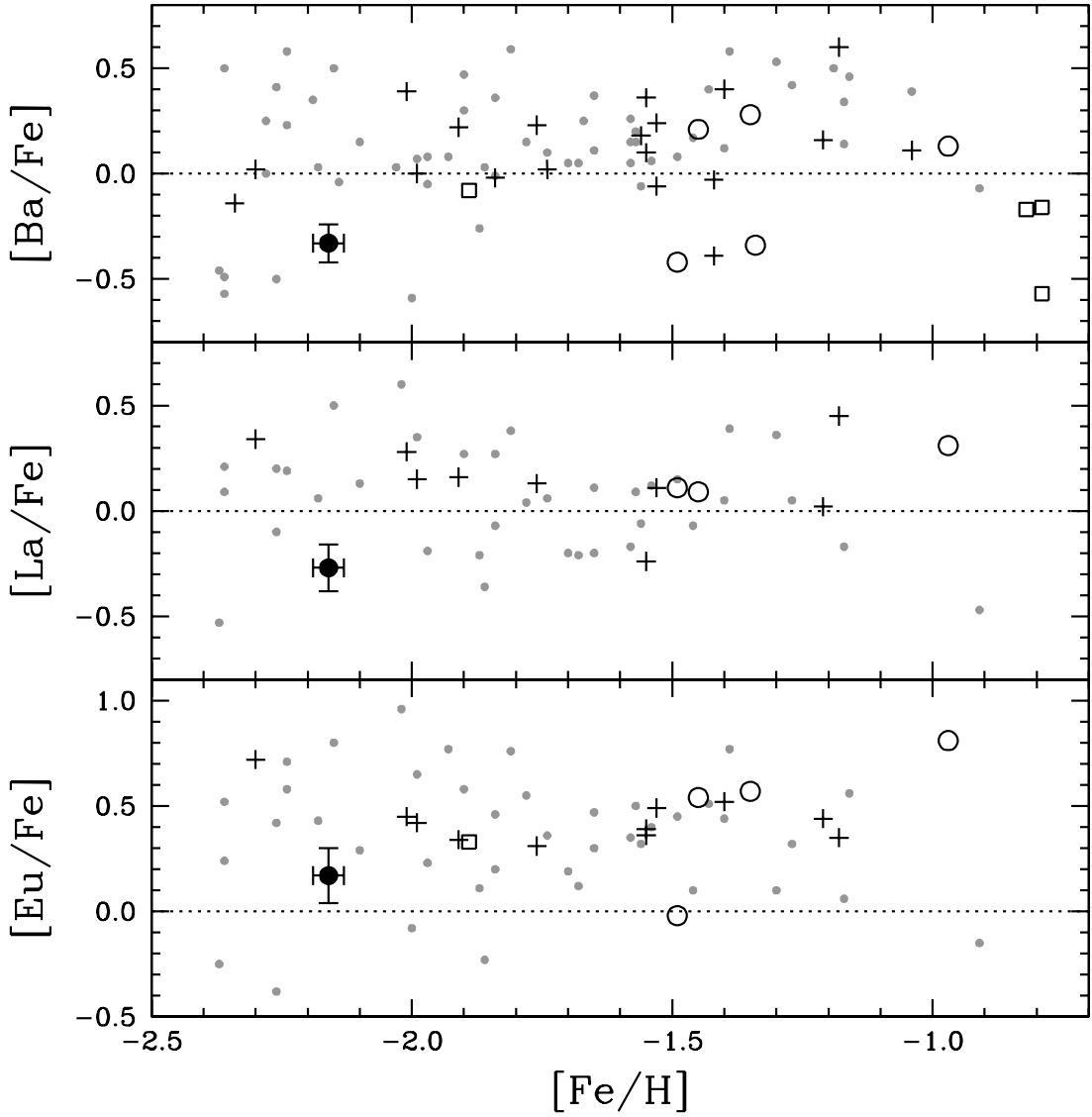}
\caption{Abundances of the neutron capture elements.
Crosses are old halo clusters, open circles are younger halo clusters,
open squares are thick disk clusters, and grey dots are field stars
(Burris et al.\ 2000).
M68 is represented by filled circles.\label{fig:neutron}}
\end{figure}

Burris et al.\ (2000) studied abundances of neutron capture
elements in a large sample of metal-poor giants, finding that a large
star-to-star variations in the neutron capture elemental abundances
(see also Gilroy et al.\ 1988, McWilliam et al.\ 1995).
They suggested that this scatter in neutron capture elemental abundances
results from inhomogeneity of the proto-stellar material polluted by SNe~II
nucleosynthesis ejecta at early stage in the Galaxy's history.
Whether this is the correct interpretation or whether the
analyses themselves are partly responsible for the scatter
remains to be seen (see Cayrel et al.\ 2004).
In Figure~\ref{fig:neutron}, we show Ba, La, and Eu abundances for
globular clusters (Brown et al.\ 1997, 1999;
Gratton et al.\ 1986; Gratton 1987; Gratton \& Ortolani 1989;
Ivans et al.\ 1999, 2001; Kraft et al.\ 1998;  Lee \& Carney 2002;
McWilliam et al.\ 1992; Shetrone \& Keane 2000;
Sneden et al.\ 1997, 2000a, 2000b, 2004)
and field stars (Burris et al.\ 2000) as a function of metallicity.
Table~\ref{tab:neutron} summarizes mean Ba, La, Eu abundances for
M68 and M15 (Sneden et al.\ 1997, 2000a), other globular clusters,
and field stars with $-$2.50 $\leq$ [Fe/H] $\leq$ $-$1.90 (Burris et al.\ 2000).

The comparison with M15 and other clusters and stars
is not quite so simple, however, as it
appears. 
Sneden et al.\ (1997) discovered that whereas the abundances
of the lighter elements silicon, calcium, and titanium did not
vary from star to star, those of the neutron capture elements did,
and by factors of four to five. Like the variation in [Si/Ti]
seen between M68 and other clusters, the variations in [Ba/Fe]
and [Eu/Fe] seen by Sneden et al.\ (1997, 2000) warn us that we
cannot assume all supernovae produce the same heavy element
abundance yields. In the case of M15, the lack of variations
in the abundances of the light elements would predict in such
a simple model that barium and other $s$-process elements
and the $r$-process element europium should likewise be constant,
contrary to what was seen.

\begin{figure}
\epsscale{1}
\figurenum{9}
\plotone{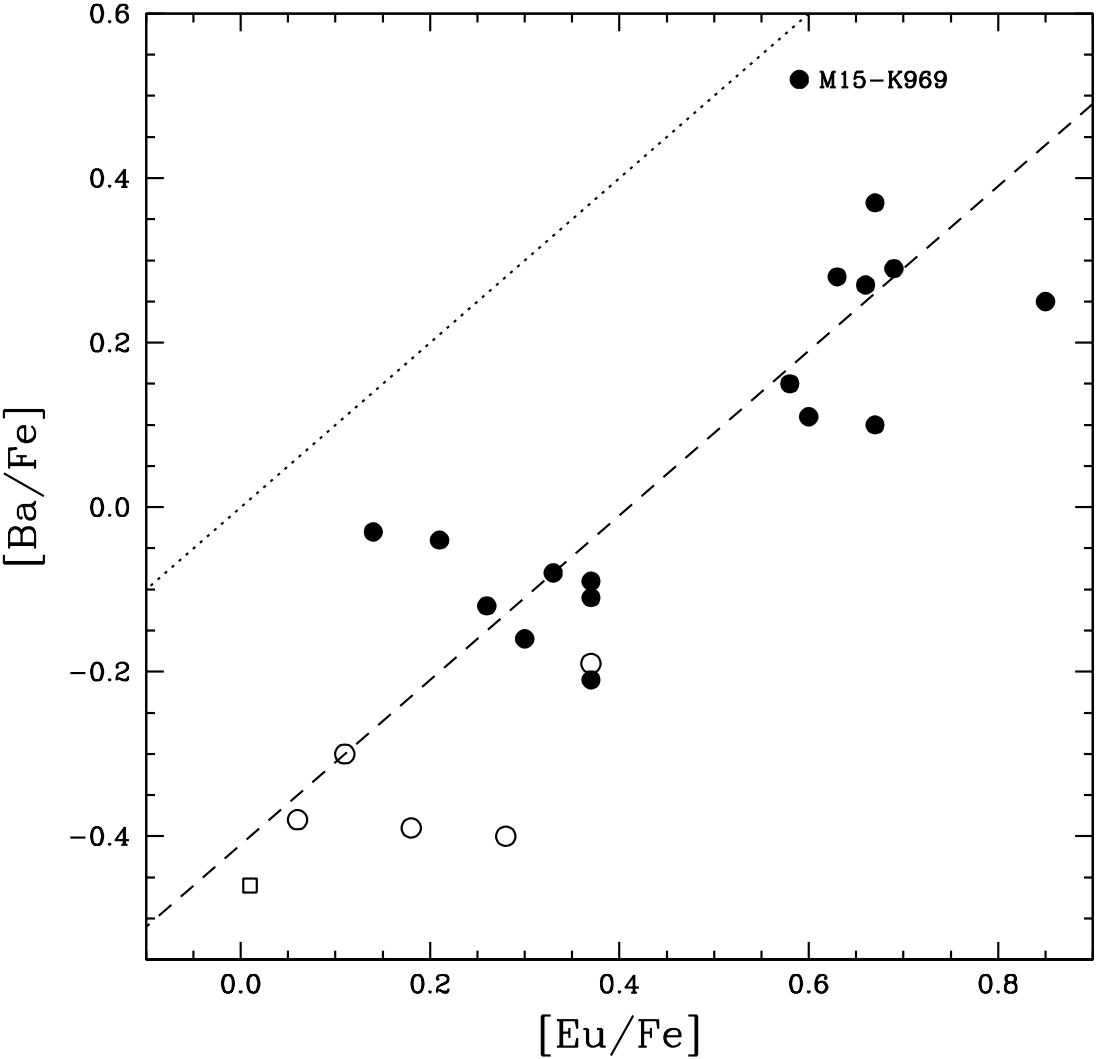}
\caption{The abundances of barium and europium relative to iron
in M15 (dots) and M68 (circles). The post-AGB star M68-117 is
marked by an open square. The dotted line represents the
solar abundance pattern. The dashed line is the approximate
mean for the stars in M15. Despite varying levels of
neutron capture abundances in the cluster, the [Ba/Eu]
ratio appears to be constant. That is true as well for M68.
\label{fig:baeu}}
\end{figure}

How does M68 fit into this picture? In Figure~\ref{fig:baeu} we
reproduce the [Ba/Fe] vs.\ [Eu/Fe] abundances found by Sneden
et al.\ (1997). The dotted line shows the solar abundances, and
the dashed line, displaced from the solar relation by 0.41 dex,
shows the mean behavior of the stars in M15, which are plotted
as dots. As Sneden et al.\ (1997) noted, the behavior of the
stars in M15 is consistent with an enhanced $r$-process
contribution to the abundances of \underline{both} barium and europium
in M15. That process produced different absolute amounts of
the neutron capture elements in stars in M15, but the process
itself yielded the same relative abundances of the two elements.
Our results for the stars in M68 are consistent with this
general picture, in that the barium and europium abundance ratio
is constant, and essentially the same in M68 and in M15.
Sneden et al.\ (1997) found
that the ``low barium" and ``high barium" groups of stars
in M15 had similar [Ba/Eu] ratios, $-0.41 \pm 0.03$ ($\sigma = 0.13$)
dex for all eighteen stars. Our seven stars, including the post-AGB
star 117, yield $<$[Ba/Eu]$>$ = $-0.43 \pm 0.05$ ($\sigma = 0.13$) dex.
What \underline{is} different between the two clusters, however,
is that both barium and europium are lower in abundance in M68
than the stars in M15, even the ``low barium" stars. Thus M68
was slightly less enriched in $r$-process nucleosynthesis,
relative to iron. We could interpret our results in
an alternative fashion, that iron is enhanced relative to the
$r$-process elements, in this case including barium and europium.

For [La/Eu] we must turn to the smaller sample of Sneden et al.\ (2000),
who studied three stars in M15 with very different [Ba/Fe] ratios,
$-0.24$, +0.05, and +0.05. The [La/Fe] values also vary greatly
for these three stars: $-0.02$, +0.38, and +0.61, yet the
variation seen in [La/Eu] is much smaller, and consistent
with the measurement uncertainties: $-0.33$, $-0.45$, and
$-0.26$, for a mean value of $-0.35 \pm 0.06$ ($\sigma = 0.10$) dex.
We were able to estimate [La/Eu] values for three
of our program stars, and we find $<$[La/Eu]$>$ = $-0.38 \pm 0.10$
($\sigma = 0.17$) dex. Thus for both lanthanum and barium, it appears
the $r$ process is dominating the nucleosynthesis enrichment in
both M68 and in M15, and yielding constant relative abundances of
the elements. But as in the case of barium and europium, there is
an underabundance of lanthanum relative to iron in M68 compared to M15.

\subsection{Summary of Differences Between M68 and M15 and Other Clusters}

\begin{figure}
\epsscale{1}
\figurenum{10}
\plotone{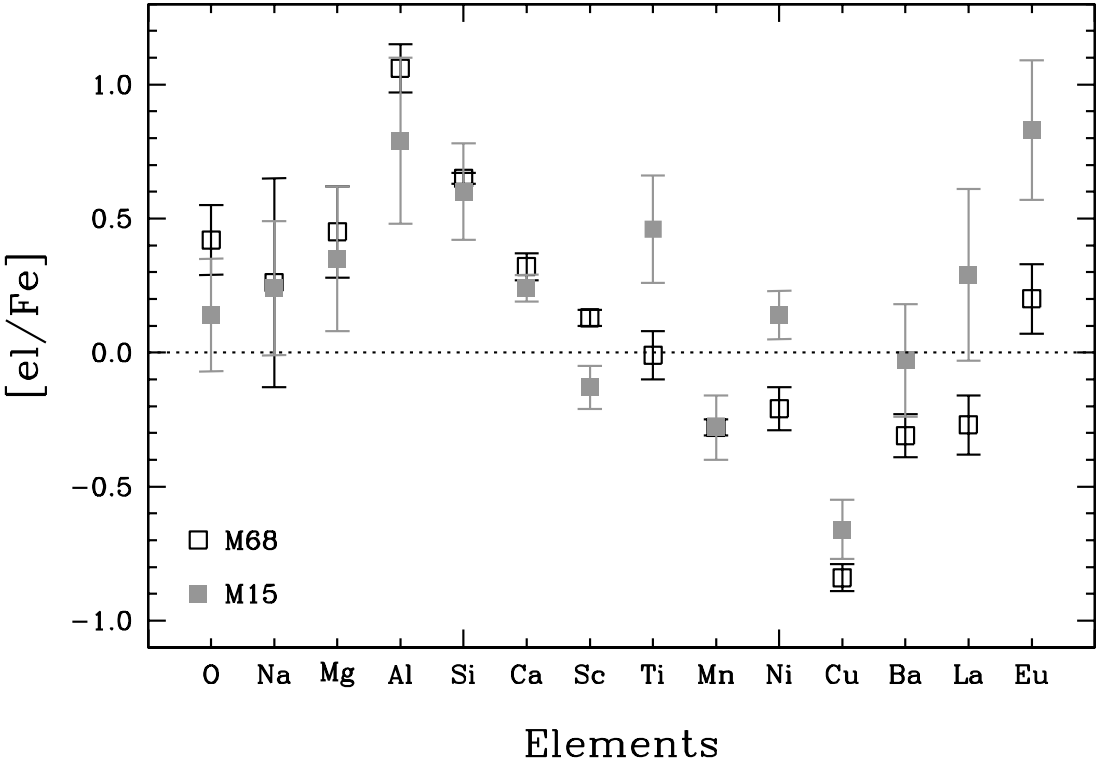}
\caption{Comparison of elemental abundances between M68 and M15.
\label{fig:compm15}}
\end{figure}

In Figure~\ref{fig:compm15}, we summarize graphically 
the many similarities between the abundances of
various elements, relative to iron, in M68 compared to
M15. Perhaps the two most striking differences are in
the much lower titanium abundances in M68, which, as we have
seen, might be explained by a greater contribution from more
massive stars and their supernovae events in M68 relative
to M15. On the other hand, while it appears that the $r$-process
has dominated the production of the ``traditional" $s$-process
elements lanthanum and barium, and that the [La/Eu] and
[Ba/Eu] ratios are the same for all the stars in both clusters,
the overall $r$-process enrichment varies within M15 (Sneden
et al.\ 1997) and between M68 and M15. The clusters have
clearly experienced somewhat different chemical enrichment
histories, and it will be interesting to see if models
of supernovae enrichment ultimately prove successful
in explaining these differences and what this means for
the histories and, perhaps, ages of these clusters.

\begin{figure}
\epsscale{1}
\figurenum{11}
\plotone{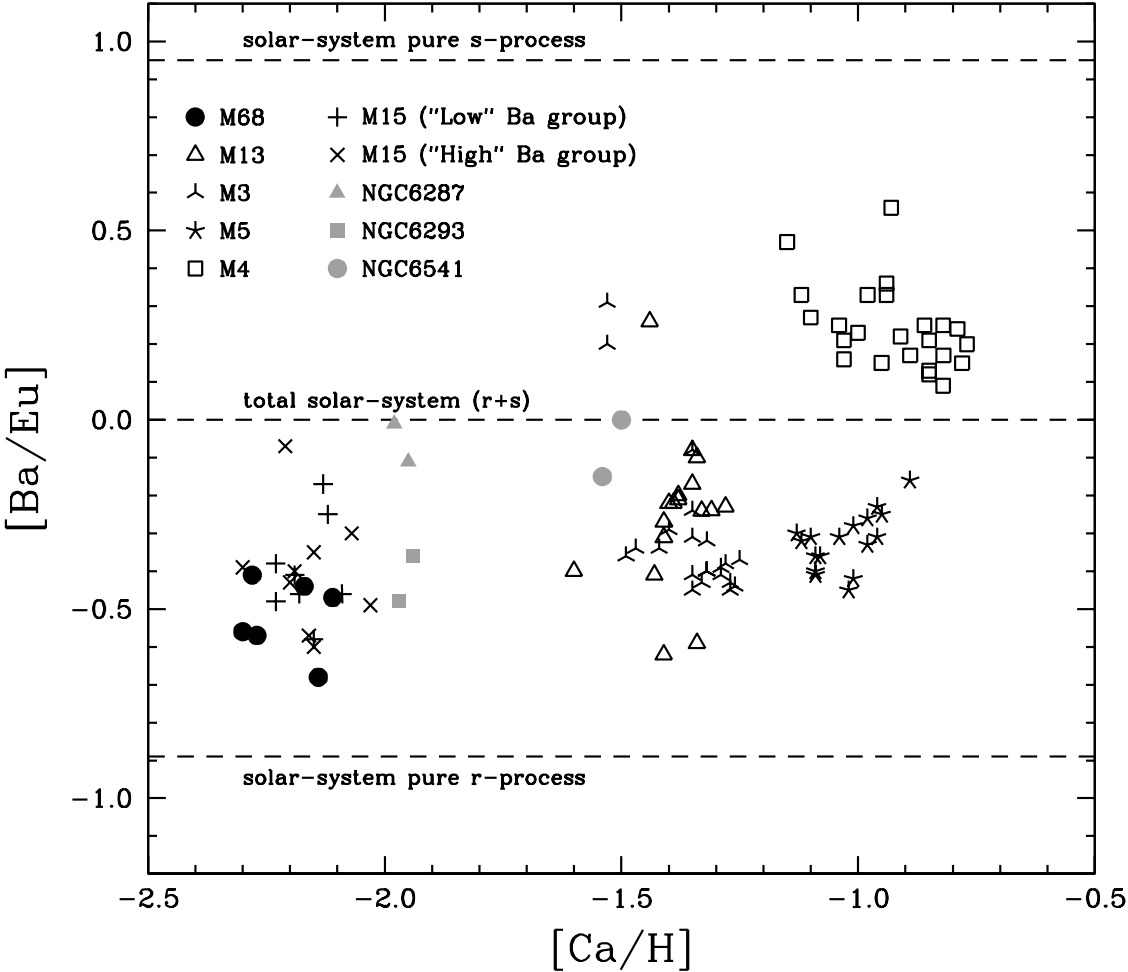}
\caption{Comparisons of [Ba/Eu] ratios as a function of [Ca/H]
of our program stars in M68 and
individual stars in other globular clusters.
\label{fig:cabaeu}}
\end{figure}

Finally, in Figure~\ref{fig:cabaeu} we show the results for
individual stars in the abundance ratio of barium and europium,
plotted against [Ca/H] rather than [Fe/H]. As Sneden et al.\ (1997)
commented, calcium is thought to be produed in Type~II supernovae,
compared to iron, whose abundance is more strongly affected by
contributions from Type~Ia supernovae. Also following
Sneden et al.\ (1997), we show the levels expected from
``pure" $r$-process nucleosynthesis, ``pure" $s$-process
nucleosynthesis, and the intermediate case found to
exist in the solar system. At lower [Ca/H] levels, 
[Ba/Eu] ratios more consistent with $r$-process domination
are apparent, which is not terribly surprising. But M68 appears
to be one of the most extreme cases. 

\section{CONCLUSIONS}

A chemical abundance study of seven giant stars in M68 has been
presented. We estimate the stars' temperatures
using the abundances derived from Fe~I lines with differing
excitation potentials, but the gravity has been derived using available
photometry rather than a comparison between abundances
from pressure-insensitive Fe~I lines and pressure-sensitive Fe~II
line. The ``spectroscopic" gravities do not agree with the
photometric gravities, suggesting non-LTE effects are
present. Using only the Fe~II lines, which should be
less sensitive to non-LTE, and photometric gravities, we find
[Fe/H] = $-$2.16 $\pm$ 0.02 ($\sigma = 0.04$. 
We have compared our results
to those of
Gratton \& Ortolani (1989), who found a higher value. We
attribute the difference to the lower-resolution spectra
available to them at the time. Our results do agree well 
with those obtained by Minniti et al.\ (1993), but we
regard the agreement as accidental since we could not
reproduce their results using their data.

We determine element-to-iron ratios using neutral vs.\
neutral and ionized vs.\ ionized lines to again minimize
non-LTE effects. We find a large range in sodium
abundances, but no significant range in oxygen abundances.
Further, the post-AGB star M68-117 does not appear to
show any enhancement of sodium. These results are not
consistent with deep mixing being the cause of the variations
among the light elements oxygen, sodium, magnesium, and
aluminum.

There are two notable differences between M68 and the comparably
metal-poor cluster M15. While both show enhanced [Si/Fe] ratios
relative to other clusters and comparably metal-poor field stars,
M68 is quite deficient in titanium compared to M15 or any
other cluster. It is possible that this arose because
nucleosynthesis enrichment of the stars in M68 was provided
by supernovae resulting from the deaths of somewhat more
massive progenitors. This would be difficult to reconcile
with the nominal younger age for M68 compared to M15, based
on the morphologies of their horizontal branch stars' distribution in
temperature/color. Perhaps age is not the only ``second parameter"?

The second interesting difference is that [La/Eu] and [Ba/Eu] ratios
are similar for the stars in M68 and in M15 (and in both its
``high barium" and ``low barium" groups of stars---Sneden et al.\ 1997).
This suggests that in both clusters the $r$ process is making a major
contribution to the abundance levels of the ``traditional"
$s$-process elements lanthanum and barium. However, whatever the
process is, it is not contributing quite as much to the neutron
capture elements' abundances in M68 as in M15.

%==========================================================================
\acknowledgments
This research was supported by the National Aeronautics and Space
Administration (NASA) grant number GO-07318.04-96A from the Space Telescope
Science Institute, which is operated by the Association of Universities
for Research in Astronomy (AURA), Inc., under NASA contract NAS 5-26555.
We also thank the National Science Foundation for financial support via
grants AST$-$9619381, AST$-$9888156 and AST$-$030541
to the University of North Carolina.
Support for this work was also provided by the Korea Science
and Engineering Foundation (KOSEF) to the Astrophysical Research Center
for the Structure and Evolution of the Cosmos (ARCSEC).

\clearpage
%==========================================================================

\clearpage

\begin{landscape}
\begin{deluxetable}{llccccccccccc}
\tablecaption{Journal of observations.\label{tab:obs}}
\tablenum{1}
\tablewidth{0pc}
\tablehead{
\multicolumn{2}{c}{Star\tablenotemark{1}} &
\multicolumn{1}{c}{} &
\multicolumn{2}{c}{Position (2000)\tablenotemark{2}} &
\multicolumn{1}{c}{$V$\tablenotemark{3}} &
\multicolumn{1}{c}{$(B-V)$\tablenotemark{3}} &
\multicolumn{1}{c}{$K$\tablenotemark{4}} &
\multicolumn{1}{c}{$K$\tablenotemark{5}} &
\multicolumn{1}{c}{Date/Time} &
\multicolumn{1}{c}{$t_{exp}$} &
\multicolumn{1}{c}{S/N} \\
\colhead{W} &
\colhead{Alt.} &
\colhead{} &
\colhead{$\alpha$} &
\colhead{$\delta$} &
\colhead{} &
\colhead{} &
\colhead{} &
\colhead{} &
\colhead{} &
\colhead{(min)} &
\colhead{}}\startdata
93  & I--82 & & 12:39:20.9 & $-$26:46:54 & 12.67 & 1.32 &  9.53 &  ~9.57 & 1996 May 4 & 120 & 170 \\
117 & ZNG2\tablenotemark{6} & & 12:39:22.5 & $-$26:45:12 & 12.69 & 1.19 &  9.73 & ~9.73 & 1996 May 6 & 120 & 115 \\
160 & I--260 & & 12:39:24.7 & $-$26:43:33 & 12.69 & 1.32 &  9.52 & ~9.53 & 1996 May 4 & 120 & 160 \\
89  & A--14  & & 12:39:20.8 & $-$26:41:39 & 12.71 & 1.33 &  9.55 & ~9.57 & 1996 May 4 & 120 & 160 \\
412 & I--144 & & 12:39:30.8 & $-$26:47:53 & 12.91 & 1.25 &  9.88 &  ~9.90 & 1996 May 6 & 170 & 140 \\
350 &        & & 12:39:29.4 & $-$26:44:23 & 13.18 & 1.15 &\nodata& 10.22 & 1996 May 5 & 190 & 140 \\
48  &        & & 12:39:15.9 & $-$26:45:15 & 13.35 & 1.10 &\nodata& 10.53 & 1996 May 7 & 130 & ~90 \\
\enddata
\tablenotetext{1}{``W" from Walker (1994). ``A" from Alcaino (1977).
``I" from Harris (1975).}
\tablenotetext{2}{Positional data from 2MASS (Cutri et al.\ 2000).}
\tablenotetext{3}{Walker (1994).}
\tablenotetext{4}{Frogel, Persson, \& Cohen (1983).}
\tablenotetext{5}{2MASS photometric data using the CIT system (Cutri et al.\ 2000).}
\tablenotetext{6}{An ultraviolet bright star (Zinn, Newell, \& Gibson 1972).}
\end{deluxetable}
\end{landscape}

\clearpage

\begin{deluxetable}{rcrrrrrrrrrr}
\tablecaption{Equivalent widths.\label{tab:ew}}
\tablenum{2}
\tablewidth{0pc}

\tablehead{
\colhead{$\lambda$ (\AA)} &
\colhead{Elem.} &
\colhead{$\chi$ (eV)} &
\colhead{$\log gf$} &
\colhead{~93} &
\colhead{117} &
\colhead{160} &
\colhead{~89} &
\colhead{412} &
\colhead{350} &
\colhead{~48} &
\colhead{Ref.}}

\startdata

6300.23 & [O I] &  0.000 & $-$9.750 &    29 &    40 &    22 &    43 &    27 &    30 &\nodata& \OI \\
6363.88 & [O I] &  0.020 &$-$10.250 &\nodata&    8  &     8 &     9 &\nodata&\nodata&\nodata& \OI \\
5682.63 & Na I  &  2.102 & $-$0.700 &   24  &\nodata&     5 &\nodata&    13 &\nodata&    19 & \NaIa \\
5688.22 & Na I  &  2.100 & $-$0.460 &   44  &    13 &     9 &     8 &    26 &    31 &\nodata& \NaIa \\
6160.75 & Na I  &  2.104 & $-$1.260 &   11  &\nodata&\nodata&\nodata&     6 &     8 &\nodata& \NaIb \\
5528.42 & Mg I  &  4.350 & $-$0.360 &   92  &   110 &   115 &   115 &   105 &    89 &   101 & \MgI \\
5711.10 & Mg I  &  4.340 & $-$1.630 &   23  &    37 &    42 &    40 &    25 &\nodata&    25 & \MgI \\
6696.03 & Al I  &  3.140 & $-$1.570 &   16  &\nodata&\nodata&\nodata&     8 &    11 &\nodata& \AlI \\
6698.67 & Al I  &  3.140 & $-$1.890 &    7  &\nodata&\nodata&\nodata&     5 &     4 &\nodata& \AlI \\
5665.56 & Si I  &  4.920 & $-$2.040 &    7  &     7 &     7 &\nodata&     7 &\nodata&\nodata& \SiIGa \\
5793.08 & Si I  &  4.930 & $-$2.060 &    7  &     8 &\nodata&\nodata&\nodata&     6 &\nodata& \SiIGa \\
6243.82 & Si I  &  5.616 & $-$1.270 &    5  &\nodata&\nodata&\nodata&     6 &\nodata&\nodata& \SiISn \\
6244.48 & Si I  &  5.616 & $-$1.270 &\nodata&\nodata&\nodata&\nodata&     5 &\nodata&\nodata& \SiISn \\
5590.11 & Ca I  &  2.521 & $-$0.710 &   32  &    34 &    37 &\nodata&    28 &    27 &    20 & \CaISn \\
5594.46 & Ca I  &  2.523 &    0.097 &   78  &    78 &    78 &\nodata&    73 &    61 &       & \CaI \\
6161.30 & Ca I  &  2.523 & $-$1.266 &   15  &    15 &    15 &    15 &    12 &     8 &\nodata& \CaI \\
6166.44 & Ca I  &  2.521 & $-$1.142 &   18  &    15 &    17 &    16 &    14 &    10 &\nodata& \CaI \\
6169.04 & Ca I  &  2.523 & $-$0.797 &   31  &    27 &    33 &    33 &    29 &    25 &    19 & \CaI \\
6169.56 & Ca I  &  2.526 & $-$0.478 &   45  &    41 &    49 &\nodata&    40 &    33 &    36 & \CaI \\
6455.60 & Ca I  &  2.523 & $-$1.290 &\nodata&    12 &    13 &    13 &\nodata&\nodata&\nodata& \CaI \\
6471.66 & Ca I  &  2.526 & $-$0.686 &   43  &    35 &    37 &    52 &\nodata&    32 &    23 & \CaI \\
6499.65 & Ca I  &  2.523 & $-$0.818 &   32  &    28 &    34 &    39 &    27 &    24 &\nodata& \CaI \\
7148.15 & Ca I  &  2.709 &    0.137 &   85  &    82 &    86 &    89 &\nodata&    62 &    65 & \CaI \\
5526.79 & Sc I  &  1.768 & $-$0.256 &    75 &    75 &    79 &    77 &    73 &    62 &    62 & \ScII \\
5657.90 & Sc I  &  1.507 & $-$0.645 &    69 &    76 &\   67 &    71 &    63 &    64 &    62 & \ScII \\
6245.64 & Sc I  &  1.507 & $-$1.134 &    37 &    35 &\nodata&    41 &    37 &    27 &    23 & \ScII \\
6604.60 & Sc I  &  1.357 & $-$1.309 &    36 &    33 &    33 &    31 &    29 &    24 &    22 & \ScII \\
5866.45 & Ti I  &  1.067 & $-$0.784 &   28  &    23 &    37 &    31 &    21 &    16 &    15 & \TiIa \\
5899.30 & Ti I  &  1.053 & $-$1.098 &   15  &    13 &\nodata&    17 &    14 &\nodata&\nodata& \TiIa \\
5922.11 & Ti I  &  1.046 & $-$1.410 &   11  &    11 &    15 &    11 &\nodata&     6 &\nodata& \TiIb \\
5953.16 & Ti I  &  1.887 & $-$0.273 &   10  &\nodata&    15 &\nodata&\nodata&\nodata&\nodata& \TiIc \\
5965.83 & Ti I  &  1.879 & $-$0.353 &    7  &\nodata&\nodata&    10 &\nodata&\nodata&\nodata& \TiIc \\
5978.54 & Ti I  &  1.870 & $-$0.500 &\nodata&    8  &     9 &\nodata&\nodata&\nodata&\nodata& \TiISn \\
6126.22 & Ti I  &  1.067 & $-$1.369 &\nodata&\nodata&\nodata&    10 &     8 &\nodata&\nodata& \TiIb \\
6258.11 & Ti I  &  1.440 & $-$0.299 &   27  &    25 &    34 &    34 &    21 &    16 &    13 & \TiIc \\
6258.71 & Ti I  &  1.460 & $-$0.270 &   35  &    25 &    40 &    36 &    24 &    20 &    15 & \TiIc \\
6261.11 & Ti I  &  1.430 & $-$0.423 &   26  &    20 &    30 &    29 &    19 &    15 &    10 & \TiIc \\
6606.95 & Ti II &  2.061 & $-$2.790 &    5  &     8 &     8 &     5 &     5 &     4 &\nodata& \TiII \\
7214.74 & Ti II &  2.590 & $-$1.740 &   11  &    14 &    13 &    10 &    12 &    10 &\nodata& \TiII \\
6021.79 & Mn I  &  3.080 &    0.030 &   21  &    22 &    22 &    21 &    17 &    16 &    10 & \MnI \\
5662.51 & Fe I  &  4.178 & $-$0.590 &   26  &    28 &    26 &    30 &    22 &    21 &   15  & \FeIOB \\
5701.55 & Fe I  &  2.559 & $-$2.216 &   56  &    52 &    57 &    55 &    49 &    41 &   39  & \FeIb \\
5753.12 & Fe I  &  4.260 & $-$0.705 &   17  &\nodata&    17 &    17 &    16 &    14 &   13  & \FeIOB \\
5816.37 & Fe I  &  4.549 & $-$0.618 &   10  &\nodata&\nodata&    10 &\nodata&\nodata&\nodata& \FeIOB \\
5916.25 & Fe I  &  2.453 & $-$2.994 &   24  &  21   &    28 &    24 &    23 &\nodata&   10  & \FeIa \\
5956.69 & Fe I  &  0.859 & $-$4.608 &   64  &  58   &\nodata&    69 &    50 &\nodata&\nodata& \FeIc \\
6027.05 & Fe I  &  4.076 & $-$1.106 &   13  &\nodata&    13 &    14 &    11 &    11 &\nodata& \FeIOB \\
6065.48 & Fe I  &  2.609 & $-$1.530 &   97  &  88   &   102 &    96 &    85 &    75 &   77  & \FeIb \\
6082.71 & Fe I  &  2.223 & $-$3.573 &\nodata&  13   &    17 &    17 &    13 &\nodata&\nodata& \FeIa \\
6151.62 & Fe I  &  2.176 & $-$3.299 &   28  &  26   &    33 &    31 &    23 &    21 &   17  & \FeIa \\
6173.34 & Fe I  &  2.223 & $-$2.880 &   48  &\nodata&    53 &    44 &    40 &    35 &   32  & \FeIa \\
6180.20 & Fe I  &  2.728 & $-$2.637 &   23  &\nodata&\nodata&    21 &\nodata&\nodata&   11  & \FeIBKb \\
6200.31 & Fe I  &  2.609 & $-$2.437 &   42  &  40   &    45 &    42 &    37 &    30 &   28  & \FeIb \\
6219.28 & Fe I  &  2.198 & $-$2.433 &   84  &  74   &    84 &    82 &    67 &    65 &   50  & \FeIa \\
6229.23 & Fe I  &  2.845 & $-$2.846 &\nodata&\nodata&    13 &    12 &\nodata&    10 &\nodata& \FeIBKa\\
6230.73 & Fe I  &  2.559 & $-$1.281 &  126  & 114   &   120 &   122 &   111 &   102 &   95  & \FeIb  \\
6232.64 & Fe I  &  3.654 & $-$1.283 &   27  &  28   &    29 &    28 &    23 &\nodata&   17  & \FeIBKb\\
6246.32 & Fe I  &  3.603 & $-$0.894 &   52  &  53   &    58 &    54 &    52 &    41 &   35  & \FeIOB \\
6252.55 & Fe I  &  2.404 & $-$1.687 &  108  & 100   &   112 &   109 &   102 &    90 &   83  & \FeIa  \\
6265.13 & Fe I  &  2.176 & $-$2.550 &   79  &  68   &    83 &    81 &    66 &    64 &   54  & \FeIa  \\
6270.22 & Fe I  &  2.858 & $-$2.505 &\nodata&  19   &\nodata&    20 &    16 &\nodata&\nodata& \FeIBKa\\
6322.69 & Fe I  &  2.588 & $-$2.426 &   41  &  45   &    51 &    45 &    41 &    36 &   28  & \FeIb  \\
6335.33 & Fe I  &  2.198 & $-$2.194 &   91  &  87   &\nodata&    99 &    81 &    77 &   67  & \FeIOB \\
6336.82 & Fe I  &  3.686 & $-$0.916 &\nodata&  41   &\nodata&\nodata&\nodata&\nodata&\nodata& \FeIBKb\\
6344.15 & Fe I  &  2.433 & $-$2.923 &   33  &  27   &    36 &    36 &    28 &    24 &  18   & \FeIa  \\
6393.60 & Fe I  &  2.433 & $-$1.469 &  117  & 108   &   118 &   116 &   105 &    99 &  92   & \FeIBKa\\
6408.02 & Fe I  &  3.686 & $-$1.066 &  35   &  34   &    40 &    41 &    29 &    31 &  23   & \FeIBKa\\
6411.65 & Fe I  &  3.654 & $-$0.734 &  59   &  59   &    65 &    61 &    54 &    43 &  42   & \FeIOB \\
6421.35 & Fe I  &  2.280 & $-$2.027 & 104   & 100   &   110 &   106 &    94 &    81 &  79   & \FeIa  \\
6430.84 & Fe I  &  2.176 & $-$2.006 & 107   & 108   &   115 &   114 &   109 &    92 &  78   & \FeIa  \\
6481.87 & Fe I  &  2.279 & $-$2.984 &  47   &  42   &    52 &\nodata&    41 &\nodata&\nodata& \FeIa  \\
6494.98 & Fe I  &  2.404 & $-$1.273 & 142   & 136   &   131 &   140 &   128 &\nodata&  107  & \FeIa  \\
6498.94 & Fe I  &  0.958 & $-$4.687 &  52   &  42   &    62 &    49 &    37 &    36 &   26  & \FeIc  \\
6574.23 & Fe I  &  0.990 & $-$5.004 &  25   &  26   &    36 &    34 &    26 &    17 &   13  & \FeIc  \\
6575.02 & Fe I  &  2.588 & $-$2.727 &  27   &  29   &    33 &    32 &    27 &    23 &   13  & \FeIOB \\
6581.21 & Fe I  &  1.485 & $-$4.707 &  13   &\nodata&    17 &    14 &\nodata&\nodata&\nodata& \FeIBKa\\
6592.91 & Fe I  &  2.728 & $-$1.490 &  98   &  88   &    96 &    94 &    87 &    79 &   68  & \FeIOB \\
6593.87 & Fe I  &  2.433 & $-$2.422 &  67   &  55   &    65 &    69 &    56 &    50 &   38  & \FeIa  \\
6609.11 & Fe I  &  2.559 & $-$2.692 &  32   &  35   &    38 &    36 &    33 &    24 &   16  & \FeIb  \\
6625.02 & Fe I  &  1.011 & $-$5.366 &  13   &\nodata&    19 &    14 &    12 &     9 &\nodata& \FeIc  \\
6663.44 & Fe I  &  2.420 & $-$2.479 &  62   &  62   &    70 &    64 &    54 &    47 &   41  & \FeIa  \\
6677.99 & Fe I  &  2.692 & $-$1.435 & 105   & 104   &   109 &   103 &    98 &    89 &   77  & \FeIOB \\
6750.15 & Fe I  &  2.424 & $-$2.621 &  51   &  50   &    62 &    60 &    50 &    42 &   34  & \FeIa  \\
6978.85 & Fe I  &  2.484 & $-$2.500 &  57   &  48   &    67 &    58 &    51 &    41 &   35  & \FeIa  \\
7223.66 & Fe I  &  3.017 & $-$2.269 &  25   &  25   &    28 &    27 &    24 &    18 &\nodata& \FeIBKb\\
7511.01 & Fe I  &  4.178 &    0.082 &  75   &  70   &    74 &    74 &    64 &    59 &   50  & \FeIOB \\
7710.36 & Fe I  &  4.220 & $-$1.129 &\nodata&\nodata&    13 &\nodata&\nodata&\nodata&\nodata& \FeIOB \\
7723.20 & Fe I  &  2.279 & $-$3.617 &  18   &\nodata&    20 &    14 &    13 &\nodata&\nodata& \FeIa  \\
5991.38 & Fe II &  3.153 & $-$3.557 &\nodata&  12   &\nodata&\nodata&\nodata&\nodata&\nodata& \FeII  \\
6149.26 & Fe II &  3.889 & $-$2.724 &    10 &  12   &\nodata&   10  &\nodata&\nodata&\nodata& \FeII  \\
6247.56 & Fe II &  3.891 & $-$2.329 &\nodata&\nodata&    21 &   25  &   19  &   20  &    19 & \FeII  \\
6416.92 & Fe II &  3.891 & $-$2.740 &\nodata&  12   &\nodata&\nodata&\nodata&\nodata&\nodata& \FeII  \\
6432.68 & Fe II &  2.891 & $-$3.708 &    20 &  22   &    19 &   21  &   19  &   18  &    15 & \FeII  \\
6456.38 & Fe II &  3.903 & $-$2.075 &    33 &  32   &    29 &   31  &   32  &   31  &    24 & \FeII  \\
6176.81 & Ni I  &  4.088 & $-$0.530 &    6  &\nodata&\nodata&\nodata&\nodata&\nodata&\nodata& \NiI   \\
6586.31 & Ni I  &  1.951 & $-$2.810 &   15  &    10 &    22 &   18  &    14 &     8 &\nodata& \NiI   \\
6643.63 & Ni I  &  1.676 & $-$2.010 &   80  &    67 &    87 &   80  &    70 &    55 &    55 & \NiISn \\
6767.77 & Ni I  &  1.826 & $-$2.170 &   71  &    57 &    74 &   66  &    58 &    44 &    34 & \NiI   \\
5782.13 & Cu I  &  1.642 & $-$1.700 &   11  &\nodata&    11 &    9  &\nodata&     6 &\nodata& \CuI   \\
5853.69 & Ba II &  0.600 & $-$1.010 &   72  &    56 &    69 &   74  &    65 &    62 &    63 & \BaII  \\
6141.73 & Ba II &  0.700 & $-$0.080 &  131  &   109 &   120 &  113  &   117 &   113 &   103 & \BaII  \\
6496.91 & Ba II &  0.600 & $-$0.380 &  127  &   109 &   119 &  122  &   119 &   113 &    98 & \BaII  \\
6390.48 & La II &  0.321 & $-$1.410 &    5  &\nodata&     5 &\nodata&     6 &\nodata&\nodata& \LaII  \\
6774.27 & La II &  0.126 & $-$1.750 &    7  &\nodata&\nodata&\nodata&     6 &\nodata&\nodata& \LaII  \\
6645.06 & Eu II &  1.380 &    0.204 &    6  &     5 &     8 &     9 &     7 &    10 &\nodata& \EuII  \\
7217.55 & Eu II &  1.230 & $-$0.301 &    3  &\nodata&     7 &     8 &\nodata&\nodata&\nodata& \EuII  \\
7426.57 & Eu II &  1.279 & $-$0.149 &    5  &\nodata&\nodata&\nodata&\nodata&\nodata&\nodata& \EuII  \\

\enddata
\end{deluxetable}

\clearpage

\begin{deluxetable}{cll}
\tablecaption{References for $\log gf$ values.\label{tab:gf}}
\tablenum{3}
\tablewidth{0pc}

\tablehead{
\colhead{Ref. No. } & \colhead{Reference} &
\colhead{Elem.}}
\startdata
\OI       & Lambert (1978)           & [O I] \\
\NaIa     & Ivans et al.\ (2001)      & Na I \\
\NaIb     & Kraft et al.\ (1992)      & Na I \\
\MgI      & Sneden et al.\ (2004)     & Mg I, Si I, Ca I, \\
          &                          & Ti I, Ni I, La II\\
\AlI      & Sneden et al.\ (1997)     & Al I, Ba II\\
\SiIGa    & Garz (1973)              & Si I \\
\CaI      & Smith (1981)             & Ca I \\
\TiIa     & Blackwell et al.\ (1982a) & Ti I \\
\TiIb     & Blackwell et al.\ (1983)  & Ti I \\
\TiIc     & Blackwell et al.\ (1986b) & Ti I \\
\TiII     & Fuhr et al.\ (1988a)      & Ti II \\
\MnI      & Prochaska \& McWilliam (2000) & Sc I, Mn I \\
\FeIa     & Blackwell et al.\ (1982b) & Fe I \\
\FeIb     & Blackwell et al.\ (1982c) & Fe I \\
\FeIc     & Blackwell et al.\ (1986a) & Fe I \\
\FeIOB    & O'Brian et al.\ (1991)    & Fe I \\
\FeIBKa   & Bard et al.\ (1991)       & Fe I \\
\FeIBKb   & Bard \& Kock (1994)      & Fe I \\
\FeII     & Bi\'emont (1991)         & Fe II \\
\NiI      & Fuhr et al.\ (1988b)      & Ni I \\
\CuI      & Kurucz \ (1993)      & Cu I \\
\EuII     & Bi\'emont et al.\ (1982)  & Eu II\\
\enddata
\end{deluxetable}

\clearpage

\begin{deluxetable}{cccccccccccc}
\tablecaption{Model atmosphere parameters.\label{tab:pam}}
\tablenum{4}
\tablewidth{0pc}
\tablehead{
\multicolumn{1}{c}{Star} &
\multicolumn{1}{c}{} &
\multicolumn{5}{c}{$T_{\rm eff}$(K)} &
\multicolumn{1}{c}{} &
\multicolumn{1}{c}{$\log g$} &
\multicolumn{1}{c}{$\log g$} &
\multicolumn{1}{c}{$v_{\rm turb}$} \\
\cline{3-7}\\
\colhead{} &
\colhead{} &
\colhead{$(B-V)$\tablenotemark{1}} &
\multicolumn{2}{c}{$(V-K)$\tablenotemark{1}} &
\colhead{(FPC83)\tablenotemark{2}} &
\colhead{(spec)} &
\colhead{} &
\colhead{(phot)} &
\colhead{(spec)} &
\colhead{(km s$^{-1}$)}\\
\cline{4-5}\\
\colhead{} &
\colhead{} &
\colhead{} &
\colhead{(FPC83)} &
\colhead{(2MASS)} &
\colhead{} &
\colhead{} &
\colhead{} &
\colhead{} &
\colhead{} &
\colhead{}
}
\startdata
~93 & & 4220 &  4215 & 4240 &  4287 &  4200 & & 0.7 & 0.1 & 1.95 \\
117 & & 4370 &  4340 & 4335 &  4501 &  4300 & & 0.8 & 0.3 & 1.80 \\
160 & & 4200 &  4200 & 4200 &  4329 &  4100 & & 0.7 & 0.0 & 1.80 \\
~89 & & 4200 &  4200 & 4215 &  4279 &  4175 & & 0.7 & 0.0 & 1.90 \\
412 & & 4275 &  4290 & 4300 &  4372 &  4200 & & 0.8 & 0.1 & 1.75 \\
350 & & 4390 &\nodata& 4340 &\nodata&  4300 & & 1.0 & 0.3 & 1.60 \\
~48 & & 4450 &\nodata& 4450 &\nodata&  4325 & & 1.1 & 0.3 & 1.75 \\
\enddata
\tablenotetext{1}{Using relations given by Alonso et al.\ (1999).}
\tablenotetext{2}{Frogel et al.\ (1983).}
\end{deluxetable}

\clearpage

\begin{deluxetable}{lrrrrrrrrr}
\tablecaption{Elemental abundances of M68 with photometric $\log g$.\label{tab:abundph}}
\tablenum{5}
\tablewidth{0pc}
\tablehead{
\colhead{} & \colhead{~93} & \colhead{117} & \colhead{160} & \colhead{~89} & \colhead{412} & \colhead{350} &\colhead{~48} &
\colhead{Mean} &
\colhead{Mean\tablenotemark{1}} }
\startdata
[Fe/H]$_{\mathrm{I}}$  & $-$2.50 & $-$2.38 & $-$2.56 & $-$2.51 & $-$2.58 & $-$2.53 & $-$2.67 & $-$2.53 & $-$2.56 \\
~~~~n                  &      43 &      38 &      42 &      45 &      42 &      35 &      35 &       7 &       6 \\
~~~~$\sigma$           &    0.05 &    0.06 &    0.05 &    0.05 &    0.06 &    0.05 &    0.06 &    0.09 &    0.06 \\

[Fe/H]$_{\mathrm{II}}$ & $-$2.17 & $-$2.16 & $-$2.12 & $-$2.14 & $-$2.16 & $-$2.15 & $-$2.23 & $-$2.16 & $-$2.16 \\
~~~~n                  &       3 &       5 &       3 &       4 &       3 &       3 &       3 &       7 &       6 \\
~~~~$\sigma$           &    0.03 &    0.05 &    0.02 &    0.06 &    0.05 &    0.02 &    0.06 &    0.03 &    0.04 \\

 & & & & & & & & & \\

[O/Fe]                 &   +0.37 &   +0.46 &   +0.23 &   +0.43 &   +0.37 &   +0.51 &   +0.59 & +0.42 & +0.42 \\
~~~~n                  &       1 &       2 &       2 &       2 &       1 &       1 &       2 &     7 &     6 \\
~~~~$\sigma$           & \nodata &    0.20 &    0.01 &    0.20 & \nodata & \nodata &    0.05 &  0.12 &  0.13 \\

[Na/Fe]                &   +0.62 & $-$0.07 & $-$0.23 & $-$0.27 &   +0.38 &   +0.55 &   +0.39 & +0.23 & +0.24 \\
~~~~n                  &       3 &       1 &       2 &       1 &       3 &       2 &       1 &     7 &     6 \\
~~~~$\sigma$           &    0.07 & \nodata &    0.00 & \nodata &    0.08 &    0.05 & \nodata &  0.40 &  0.39 \\

[Mg/Fe]                &   +0.12 &   +0.39 &   +0.50 &   +0.47 &   +0.35 &   +0.19 &   +0.45 & +0.35 & +0.35 \\
~~~~n                  &       2 &       2 &       2 &       2 &       2 &       1 &       2 &     7 &     6 \\
~~~~$\sigma$           &    0.05 &    0.01 &    0.01 &    0.01 &    0.10 & \nodata &    0.03 &  0.14 &  0.16 \\

[Al/Fe]                &   +1.18 & \nodata & \nodata & \nodata &   +1.00 &   +1.07 & \nodata & +1.08 & +1.08 \\
~~~~n                  &       2 & \nodata & \nodata & \nodata &       2 &       2 & \nodata &     3 &     3 \\
~~~~$\sigma$           &    0.05 & \nodata & \nodata & \nodata &    0.07 &    0.07 & \nodata &  0.09 &  0.09 \\

[Si/Fe]                &   +0.68 &   +0.65 &   +0.77 & \nodata &   +0.75 &   +0.72 & \nodata & +0.71 & +0.73 \\
~~~~n                  &       3 &       2 &       1 & \nodata &       3 &       1 & \nodata &     5 &     4 \\
~~~~$\sigma$           &    0.05 &    0.04 & \nodata & \nodata &    0.05 & \nodata & \nodata &  0.05 &  0.04 \\

[Ca/Fe]                &   +0.33 &   +0.27 &   +0.29 &   +0.37 &   +0.30 &   +0.23 &   +0.32 & +0.30 & +0.31 \\
~~~~n                  &       9 &      10 &      10 &       7 &       7 &       9 &       5 &     7 &     6 \\
~~~~$\sigma$           &    0.05 &    0.07 &    0.04 &    0.08 &    0.04 &    0.07 &    0.05 &  0.04 &  0.05 \\

[Sc/Fe]$_{\mathrm{II}}$&   +0.14 &   +0.19 &   +0.16 &   +0.14 &   +0.13 &   +0.07 &   +0.13 & +0.14 & +0.13 \\
~~~~n                  &       4 &       4 &       3 &       4 &       4 &       4 &       4 &     7 &     6 \\
~~~~$\sigma$           &    0.14 &    0.20 &    0.25 &    0.20 &    0.18 &    0.18 &    0.18 &  0.04 &  0.03 \\

[Ti/Fe]$_{\mathrm{I}}$ &   +0.06 &   +0.07 &   +0.09 &   +0.03 & $-$0.02 & $-$0.02 &   +0.01 & +0.03 & +0.02 \\
~~~~n                  &       8 &       6 &       7 &       8 &       6 &       5 &       4 &     7 &     6 \\
~~~~$\sigma$           &    0.07 &    0.05 &    0.07 &    0.05 &    0.04 &    0.06 &    0.05 &  0.04 &  0.05 \\

[Ti/Fe]$_{\mathrm{II}}$& $-$0.12 &   +0.09 &   +0.01 & $-$0.15 & $-$0.06 & $-$0.08 & \nodata & $-$0.05 & $-$0.08 \\
~~~~n                  &       2 &       2 &       2 &       2 &       2 &       2 & \nodata &       6 &       5 \\
~~~~$\sigma$           &    0.03 &    0.12 &    0.08 &    0.05 &    0.01 &    0.02 & \nodata &    0.09 &    0.06 \\

[Ti/Fe]$_{\mathrm{mean}}$& $-$0.03 &   +0.08 &   +0.05 & $-$0.06 & $-$0.04 & $-$0.05 &   +0.01 & $-$0.01 & $-$0.03 \\

[Mn/Fe]                & $-$0.25 & $-$0.22 & $-$0.30 & $-$0.27 & $-$0.29 & $-$0.26 & $-$0.33 & $-$0.27 & $-$0.28 \\
~~~~n                  &       1 &       1 &       1 &       1 &       1 &       1 &       1 &       7 &       6 \\
~~~~$\sigma$           & \nodata & \nodata & \nodata & \nodata & \nodata & \nodata & \nodata &    0.04 &    0.03 \\

[Ni/Fe]                & $-$0.04 & $-$0.21 & $-$0.01 & $-$0.09 & $-$0.09 & $-$0.22 & $-$0.14 & $-$0.11 & $-$0.10 \\
~~~~n                  &       4 &       3 &       3 &       3 &       3 &       3 &       2 &       7 &       6 \\
~~~~$\sigma$           &    0.11 &    0.11 &    0.08 &    0.09 &    0.09 &    0.09 &    0.00 &    0.08 &    0.07 \\

[Cu/Fe]                & $-$0.74 & \nodata & $-$0.83 & $-$0.86 & \nodata & $-$0.85 & \nodata & $-$0.82 & $-$0.82 \\
~~~~n                  &       1 & \nodata &       1 &       1 & \nodata &       1 & \nodata &       4 &       4 \\
~~~~$\sigma$           & \nodata & \nodata & \nodata & \nodata & \nodata & \nodata & \nodata &    0.05 &    0.05 \\

[Ba/Fe]$_{\mathrm{II}}$& $-$0.28 & $-$0.46 & $-$0.39 & $-$0.40 & $-$0.30 & $-$0.19 & $-$0.32 & $-$0.33 & $-$0.31 \\
~~~~n                  &       3 &       3 &       3 &       3 &       3 &       3 &       3 &       7 &       6 \\
~~~~$\sigma$           &    0.09 &    0.12 &    0.06 &    0.10 &    0.12 &    0.14 &    0.04 &    0.09 &    0.08 \\

[La/Fe]$_{\mathrm{II}}$& $-$0.23 & \nodata & $-$0.40 & \nodata & $-$0.19 & \nodata & \nodata & $-$0.27 & $-$0.27 \\
~~~~n                  &       2 & \nodata &       1 & \nodata &       2 & \nodata & \nodata &       3 &       3 \\
~~~~$\sigma$           &    0.14 & \nodata & \nodata & \nodata &    0.03 & \nodata & \nodata &    0.11 &    0.11 \\

[Eu/Fe]$_{\mathrm{II}}$&   +0.06 &   +0.01 &   +0.18 &   +0.28 &   +0.11 &   +0.37 & \nodata & +0.17 & +0.20 \\
~~~~n                  &       3 &       1 &       2 &       2 &       1 &       1 & \nodata &     6 &     5 \\
~~~~$\sigma$           &    0.08 & \nodata &    0.13 &    0.12 & \nodata & \nodata & \nodata &  0.14 &  0.13 \\
\enddata
\tablenotetext{1}{Without the star 117.}
\end{deluxetable}

\clearpage

\begin{deluxetable}{lrrrrrrrrr}
\tablecaption{Elemental abundances of M68 spectroscopic $\log g$.\label{tab:abundsp}}
\tablenum{6}
\tablewidth{0pc}
\tablehead{
\colhead{} & \colhead{~93} & \colhead{117} & \colhead{160} & \colhead{~89} & \colhead{412} & \colhead{350} &\colhead{~48} &
\colhead{Mean} &
\colhead{Mean\tablenotemark{1}} }
\startdata
[Fe/H]$_{\mathrm{I}}$  & $-$2.40 & $-$2.30 & $-$2.46 & $-$2.40 & $-$2.47 & $-$2.43 & $-$2.56 & $-$2.43 & $-$2.45 \\
~~~~n                  &      43 &      38 &      42 &      45 &      42 &      35 &      35 &       7 &       6 \\
~~~~$\sigma$           &    0.05 &    0.06 &    0.05 &    0.06 &    0.06 &    0.05 &    0.07 &    0.08 &    0.06 \\

[Fe/H]$_{\mathrm{II}}$ & $-$2.37 & $-$2.33 & $-$2.35 & $-$2.36 & $-$2.39 & $-$2.39 & $-$2.52 & $-$2.39 & $-$2.40 \\
~~~~n                  &       3 &       5 &       3 &       4 &       3 &       3 &       3 &       7 &       6 \\
~~~~$\sigma$           &    0.02 &    0.05 &    0.02 &    0.06 &    0.04 &    0.02 &    0.06 &    0.06 &    0.06 \\

[Fe/H]$_{\mathrm{mean}}$ & $-$2.39 & $-$2.32 & $-$2.41 & $-$2.38 & $-$2.43 & $-$2.41 & $-$2.54 & $-$2.43 & $-$2.41 \\

 & & & & & & & & & \\

[O/Fe]                 &   +0.38 &   +0.43 &   +0.23 &   +0.42 &   +0.39 &   +0.51 &   +0.60 & +0.42 & +0.42 \\
~~~~n                  &       1 &       2 &       2 &       2 &       1 &       1 &       2 &     7 &     6 \\
~~~~$\sigma$           & \nodata &    0.20 &    0.01 &    0.21 & \nodata & \nodata &    0.06 &  0.11 &  0.13 \\

[Na/Fe]                &   +0.65 & $-$0.04 & $-$0.21 & $-$0.24 &   +0.38 &   +0.56 &   +0.40 & +0.21 & +0.26 \\
~~~~n                  &       3 &       1 &       2 &       1 &       3 &       2 &       1 &     7 &     6 \\
~~~~$\sigma$           &    0.08 & \nodata &    0.00 & \nodata &    0.07 &    0.05 & \nodata &  0.37 &  0.39 \\

[Mg/Fe]                &   +0.19 &   +0.48 &   +0.60 &   +0.60 &   +0.43 &   +0.32 &   +0.53 & +0.45 & +0.45 \\
~~~~n                  &       2 &       2 &       2 &       2 &       2 &       1 &       2 &     7 &     6 \\
~~~~$\sigma$           &    0.03 &    0.06 &    0.08 &    0.08 &    0.21 & \nodata &    0.13 &  0.15 &  0.17 \\

[Al/Fe]                &   +1.16 & \nodata & \nodata & \nodata &   +0.98 &   +1.04 & \nodata & +1.06 & +1.06 \\
~~~~n                  &       2 & \nodata & \nodata & \nodata &       2 &       2 & \nodata &     3 &     3 \\
~~~~$\sigma$           &    0.06 & \nodata & \nodata & \nodata &    0.07 &    0.07 & \nodata &  0.09 &  0.09 \\

[Si/Fe]                &   +0.62 &   +0.63 &   +0.67 & \nodata &   +0.66 &   +0.64 & \nodata & +0.64 & +0.65 \\
~~~~n                  &       3 &       2 &       1 & \nodata &       3 &       1 & \nodata &     5 &     4 \\
~~~~$\sigma$           &    0.06 &    0.04 & \nodata & \nodata &    0.06 & \nodata & \nodata &  0.02 &  0.02 \\

[Ca/Fe]                &   +0.36 &   +0.31 &   +0.31 &   +0.39 &   +0.31 &   +0.24 &   +0.32 & +0.32 & +0.32 \\
~~~~n                  &       9 &      10 &      10 &       7 &       7 &       9 &       5 &     7 &     6 \\
~~~~$\sigma$           &    0.04 &    0.08 &    0.07 &    0.09 &    0.06 &    0.07 &    0.05 &  0.05 &  0.05 \\

[Sc/Fe]$_{\mathrm{II}}$&   +0.23 &   +0.22 &   +0.33 &   +0.23 &   +0.24 &   +0.14 &   +0.20 & +0.23 & +0.23 \\
~~~~n                  &       4 &       4 &       3 &       4 &       4 &       4 &       4 &     7 &     6 \\
~~~~$\sigma$           &    0.19 &    0.22 &    0.34 &    0.26 &    0.24 &    0.22 &    0.22 &  0.06 &  0.06 \\

[Ti/Fe]$_{\mathrm{I}}$ &   +0.06 &   +0.08 &   +0.06 &   +0.08 & $-$0.04 & $-$0.02 &   +0.01 & +0.03 & +0.03 \\
~~~~n                  &       8 &       6 &       7 &       8 &       6 &       5 &       4 &     7 &     6 \\
~~~~$\sigma$           &    0.07 &    0.05 &    0.09 &    0.06 &    0.04 &    0.06 &    0.06 &  0.05 &  0.05 \\

[Ti/Fe]$_{\mathrm{II}}$& $-$0.10 &   +0.08 &   +0.06 & $-$0.15 & $-$0.02 & $-$0.06 & \nodata & $-$0.03 & $-$0.05 \\
~~~~n                  &       2 &       2 &       2 &       2 &       2 &       2 & \nodata &       6 &       5 \\
~~~~$\sigma$           &    0.02 &    0.13 &    0.10 &    0.06 &    0.00 &    0.03 & \nodata &    0.09 &    0.08 \\

[Ti/Fe]$_{\mathrm{mean}}$& $-$0.02 &   +0.08 &   +0.06 & $-$0.04 & $-$0.03 & $-$0.04 & +0.01 & +0.00 & $-$0.01 \\

[Mn/Fe]                & $-$0.24 & $-$0.19 & $-$0.31 & $-$0.25 & $-$0.30 & $-$0.26 & $-$0.33 & $-$0.27 & $-$0.28 \\
~~~~n                  &       1 &       1 &       1 &       1 &       1 &       1 &       1 &       7 &       6 \\
~~~~$\sigma$           & \nodata & \nodata & \nodata & \nodata & \nodata & \nodata & \nodata &    0.05 &    0.04 \\

[Ni/Fe]                & $-$0.04 & $-$0.22 & $-$0.13 & $-$0.16 & $-$0.18 & $-$0.27 & $-$0.19 & $-$0.17 & $-$0.16 \\
~~~~n                  &       4 &       3 &       3 &       3 &       3 &       3 &       2 &       7 &       6 \\
~~~~$\sigma$           &    0.11 &    0.11 &    0.09 &    0.09 &    0.09 &    0.09 &    0.00 &    0.07 &    0.08 \\

[Cu/Fe]                & $-$0.76 & \nodata & $-$0.90 & $-$0.88 & \nodata & $-$0.87 & \nodata & $-$0.86 & $-$0.86 \\
~~~~n                  &       1 & \nodata &       1 &       1 & \nodata &       1 & \nodata &       4 &       4 \\
~~~~$\sigma$           & \nodata & \nodata & \nodata & \nodata & \nodata & \nodata & \nodata &    0.06 &    0.06 \\

[Ba/Fe]$_{\mathrm{II}}$& $-$0.18 & $-$0.44 & $-$0.24 & $-$0.31 & $-$0.18 & $-$0.12 & $-$0.24 & $-$0.24 & $-$0.21 \\
~~~~n                  &       3 &       3 &       3 &       3 &       3 &       3 &       3 &       7 &       6 \\
~~~~$\sigma$           &    0.11 &    0.11 &    0.05 &    0.10 &    0.12 &    0.13 &    0.06 &    0.11 &    0.07 \\

[La/Fe]$_{\mathrm{II}}$& $-$0.20 & \nodata & $-$0.33 & \nodata & $-$0.15 & \nodata & \nodata & $-$0.23 & $-$0.23 \\
~~~~n                  &       2 & \nodata &       1 & \nodata &       2 & \nodata & \nodata &       3 &       3 \\
~~~~$\sigma$           &    0.13 & \nodata & \nodata & \nodata &    0.02 & \nodata & \nodata &    0.10 &    0.10 \\

[Eu/Fe]$_{\mathrm{II}}$&   +0.06 &   +0.01 &   +0.24 &   +0.28 &   +0.15 &   +0.39 & \nodata & +0.18 & +0.22 \\
~~~~n                  &       3 &       1 &       2 &       2 &       1 &       1 & \nodata &     6 &     5 \\
~~~~$\sigma$           &    0.08 & \nodata &    0.12 &    0.11 & \nodata & \nodata & \nodata &  0.14 &  0.13 \\
\enddata
\tablenotetext{1}{Without the star 117.}
\end{deluxetable}

\clearpage

\begin{deluxetable}{lc}
\tablecaption{Differences in elemental abundances using photometric
and spectroscopic surface gravities.\label{tab:diff}}
\tablenum{7}
\tablewidth{0pc}
\tablehead{
\colhead{Elem.} &
\colhead{[el/Fe]$_{\mathrm{phot}}$ $-$ [el/Fe]$_{\mathrm{spec}}$}}
\startdata

[Fe/H]                 & +0.25 $\pm$ 0.05 \\

[O/Fe]                 & +0.01 $\pm$ 0.01 \\

[Na/Fe]                & $-$0.02 $\pm$ 0.01 \\

[Mg/Fe]                & $-$0.10 $\pm$ 0.02 \\

[Al/Fe]                & +0.02 $\pm$ 0.01 \\

[Si/Fe]                & +0.07 $\pm$ 0.03 \\

[Ca/Fe]                & $-$0.02 $\pm$ 0.01 \\

[Sc/Fe]$_{\mathrm{II}}$& $-$0.09 $\pm$ 0.04 \\

[Ti/Fe]$_{\mathrm{I}}$ & +0.00 $\pm$ 0.03 \\

[Ti/Fe]$_{\mathrm{II}}$& $-$0.02 $\pm$ 0.02 \\

[Ti/Fe]$_{\mathrm{mean}}$& $-$0.01 $\pm$ 0.01 \\

[Mn/Fe]                & $-$0.01 $\pm$ 0.02 \\

[Ni/Fe]                & +0.05 $\pm$ 0.04 \\

[Cu/Fe]                & +0.03 $\pm$ 0.03 \\

[Ba/Fe]$_{\mathrm{II}}$& $-$0.10 $\pm$ 0.04 \\

[La/Fe]$_{\mathrm{II}}$& $-$0.05 $\pm$ 0.02 \\

[Eu/Fe]$_{\mathrm{II}}$& $-$0.02 $\pm$ 0.03 \\

\enddata
\end{deluxetable}

\clearpage

\begin{deluxetable}{lccc}
\tablecaption{Abundance dependencies on model atmosphere.\label{tab:moddep}}
\tablenum{8}
\tablewidth{0pc}
\tablehead{
\multicolumn{1}{c}{Elem.} &
\multicolumn{1}{c}{$\delta T_{\rm eff}$} &
\multicolumn{1}{c}{$\delta \log g$} &
\multicolumn{1}{c}{$\delta v_{\rm turb}$} \\

\multicolumn{1}{c}{} &
\multicolumn{1}{c}{$\pm$ 80} &
\multicolumn{1}{c}{$\pm$ 0.3} &
\multicolumn{1}{c}{$\pm$ 0.2} \\

\multicolumn{1}{c}{} &
\multicolumn{1}{c}{(K)} &
\multicolumn{1}{c}{} &
\multicolumn{1}{c}{(km s$^{-1}$)} }
\startdata

[Fe/H]$_{\mathrm{I}}$  & $\pm$ 0.12  & $\mp$ 0.03  & $\mp$ 0.03 \\

[Fe/H]$_{\mathrm{II}}$ & $\mp$ 0.06  & $\pm$ 0.10  & $\pm$ 0.02 \\

 & & & \\

[O/Fe]                 & $\pm$ 0.04  & $\pm$ 0.05  & $\mp$ 0.03 \\

[Na/Fe]                & $\pm$ 0.03  & $\mp$ 0.03  & $\pm$ 0.03 \\

[Mg/Fe]                & $\mp$ 0.05  & $\mp$ 0.05  & $\pm$ 0.01 \\

[Al/Fe]                & $\mp$ 0.06  & $\mp$ 0.01  & $\pm$ 0.03 \\

[Si/Fe]                & $\mp$ 0.10  & $\pm$ 0.01  & $\pm$ 0.04 \\

[Ca/Fe]                & $\mp$ 0.04  & $\mp$ 0.04  & $\pm$ 0.01 \\

[Sc/Fe]$_{\mathrm{II}}$& $\pm$ 0.06  & $\mp$ 0.02  & $\mp$ 0.01 \\

[Ti/Fe]$_{\mathrm{I}}$ & $\pm$ 0.03  & $\mp$ 0.02  & $\pm$ 0.03 \\

[Ti/Fe]$_{\mathrm{II}}$& $\pm$ 0.04  & $\pm$ 0.00  & $\pm$ 0.02 \\

[Mn/Fe]                & $\mp$ 0.02  & $\mp$ 0.03  & $\pm$ 0.04 \\

[Ni/Fe]                & $\mp$ 0.02  & $\pm$ 0.01  & $\pm$ 0.01 \\

[Cu/Fe]                & $\pm$ 0.01  & $\pm$ 0.01  & $\pm$ 0.02 \\

[Ba/Fe]$_{\mathrm{II}}$& $\pm$ 0.08  & $\mp$ 0.02  & $\mp$ 0.09 \\

[La/Fe]$_{\mathrm{II}}$& $\pm$ 0.09  & $\mp$ 0.01  & $\pm$ 0.02 \\

[Eu/Fe]$_{\mathrm{II}}$& $\pm$ 0.06  & $\pm$ 0.00  & $\pm$ 0.02 \\

\enddata
\end{deluxetable}

\clearpage

\begin{deluxetable}{cccc}
\tablecaption{Stellar parameters and elemental abundances
of the star 160.\label{tab:comp160}}
\tablenum{9}
\tablewidth{0pc}
\tablehead{
\colhead{} &
\colhead{This study} &
\colhead{Gratton \& Ortolani} &
\colhead{Minniti et al.}\\
\colhead{} &
\colhead{} &
\colhead{(1989)} &
\colhead{(1993 \& 1996)}}
\startdata
$T_{\rm eff}$  & 4100 & 4329 & 4400 \\
(K) & & & \\
$\log g$   &  0.7 & 0.75 &  1.0 \\
method   &  phot. & phot. &  spec. \\
$v_{\rm turb}$ &  1.8 &  1.6 &  2.0 \\
(km sec$^{-1}$) & & & \\

[Fe/H]$_{\mathrm{I}}$  & $-$2.56 & $-$1.94 & $-$2.11 \\

[Fe/H]$_{\mathrm{II}}$ & $-$2.17 & \nodata & $-$2.14 \\

[O/Fe]  &   +0.23 &   +0.22 &   +0.25 \\

[Na/Fe] & $-$0.23 & $-$0.04 & $-$0.08 \\

[Mg/Fe] &   +0.50 &   +0.07 & \nodata \\

[Si/Fe] &   +0.77 &   +0.41 & \nodata \\

[Ca/Fe] &   +0.29 &   +0.25 & \nodata \\

[Ti/Fe] &   +0.05 &   +0.43 & \nodata \\

[Ni/Fe] & $-$0.01 & $-$0.15 & \nodata \\

[Ba/Fe] & $-$0.39 & $-$0.33 & \nodata \\
\enddata
\end{deluxetable}

\clearpage

\begin{deluxetable}{ccccccc}
\tablecaption{Comparisons of $\alpha$-element abundances.\label{tab:alpha}}
\tablenum{10}
\tablewidth{0pc}

\tablehead{
\colhead{} &
\colhead{[Si/Fe]} &
\colhead{[Ca/Fe]} &
\colhead{[Ti/Fe]} &
\colhead{[$\alpha$/Fe]} &
\colhead{n\tablenotemark{1}}}
\startdata
M68 & 0.73 $\pm$ 0.02 & 0.31 $\pm$ 0.02 & $-$0.03 $\pm$ 0.04 &
0.34 $\pm$ 0.22 & 5,7,7 \\

&&&&&&\\
M15\tablenotemark{2} & 0.62 $\pm$ 0.06 & 0.24 $\pm$ 0.01 & 0.27 $\pm$ 0.08 &
0.38 $\pm$ 0.12 & 11,18,4 \\

&&&&&&\\

old halo   & 0.38 $\pm$ 0.05 & 0.30 $\pm$ 0.03 & ~~0.33 $\pm$ 0.03 &
0.34 $\pm$ 0.02 & 15 \\

old inner halo\tablenotemark{3} &
0.56 $\pm$ 0.05 & 0.26 $\pm$ 0.07 & ~~0.15 $\pm$ 0.06 & 0.32 $\pm$ 0.03 & 3 \\

younger halo & 0.29 $\pm$ 0.09 & 0.14 $\pm$ 0.06 & ~~0.17 $\pm$ 0.07 &
0.20 $\pm$ 0.05 & 7 \\

younger halo\tablenotemark{4}
& 0.38 $\pm$ 0.09 & 0.21 $\pm$ 0.04 & ~~0.25 $\pm$ 0.06 &
0.28 $\pm$ 0.02 & 5 \\

disk       & 0.39 $\pm$ 0.08 & 0.16 $\pm$ 0.07 & ~~0.31 $\pm$ 0.08 &
0.29 $\pm$ 0.04 & 4 \\

&&&&&&\\

Field\tablenotemark{5} & 0.52 $\pm$ 0.04 & 0.35 $\pm$ 0.03 & ~~0.30 $\pm$ 0.04 &
0.39 $\pm$ 0.02 & 9, 20, 20 \\
\enddata
\tablenotetext{1}{Number of stars (M68; M15; Field) or clusters.}
\tablenotetext{2}{We used [Si/Fe] and [Ca/Fe] from Sneden et al.\ (1997). For [Ti/Fe],
we also employed data from Sneden et al.\ (2000). We did not include any stars
with uncertain measures (marked with a colon in their results.}
\tablenotetext{3}{NGC~6287, NGC~6293, \& NGC~6541 (Lee \& Carney 2002).}
\tablenotetext{4}{Without Palomar~12 and Rupercht~106.}
\tablenotetext{5}{Mean abundances for field stars with
$-$2.50 $\leq$ [Fe/H] $\leq$ $-$1.90.}
\end{deluxetable}

\clearpage

\begin{deluxetable}{ccccccc}
\tablecaption{Comparisons of neutron capture element abundances.
\label{tab:neutron}}
\tablenum{11}
\tablewidth{0pc}
\tablehead{
\colhead{} &
\colhead{[Ba/Fe]} &
\colhead{n} &
\colhead{[La/Fe]} &
\colhead{n} &
\colhead{[Eu/Fe]} &
\colhead{n}\tablenotemark{1}}
\startdata
M68 & $-$0.31 $\pm$ 0.08 & & $-$0.27 $\pm$ 0.11 & & +0.20 $\pm$ 0.13 & 7,3,6\\

&&&&&&\\

M15-all\tablenotemark{2} &
+0.10 $\pm$ 0.21 & & \nodata & & +0.49 $\pm$ 0.20 & 18,0,18 \\

M15-high Ba\tablenotemark{3} &
+0.23 $\pm$ 0.09 & & \nodata & & +0.66 $\pm$ 0.08 & 10,0,10 \\
M15-low Ba\tablenotemark{4} &
$-$0.11 $\pm$ 0.06 & & \nodata & & +0.29 $\pm$ 0.09 & 8,0,8 \\

&&&&&&\\

old halo &
+0.11 $\pm$ 0.23 & 16 & +0.22 $\pm$ 0.18 & 5 & +0.45 $\pm$ 0.13 & 8 \\

old inner halo\tablenotemark{5} &
+0.21 $\pm$ 0.20 & 3 & +0.18 $\pm$ 0.08 & 3 & +0.39 $\pm$ 0.07 & 3 \\

younger halo &
$-$0.03 $\pm$ 0.33 & 5 & +0.17 $\pm$ 0.12 & 3 & +0.48 $\pm$ 0.35 & 4 \\

younger halo\tablenotemark{6}
& +0.05 $\pm$ 0.34 & 3 & +0.09~~~~~~~~~~ & 1 & +0.56 $\pm$ 0.02 & 2 \\

disk   & $-$0.25 $\pm$ 0.22 & 4 & \nodata & \nodata & +0.33~~~~~~~~~~ & 1 \\
&&&&&&\\
Field\tablenotemark{7} &
+0.18 $\pm$ 0.11 & 27 & +0.11 $\pm$ 0.08 & 14 & +0.39 $\pm$ 0.09 & 17 \\

\enddata
\tablenotetext{1}{Number of stars (M68; M15; Field) or clusters.}
\tablenotetext{2}{Sneden et al.\ (1997).}
\tablenotetext{3}{``High Ba" group (Sneden et al.\ 1997). See Figure~\ref{fig:baeu}}
\tablenotetext{4}{``Low Ba" group (Sneden et al.\ 1997). See Figure~\ref{fig:baeu}}
\tablenotetext{5}{NGC~6287, NGC~6293, \& NGC~6541 (Lee \& Carney 2002).}
\tablenotetext{6}{Without Palomar~12 and Rupercht~106.}
\tablenotetext{7}{Mean abundances for field stars with
$-$2.50 $\leq$ [Fe/H] $\leq$ $-$1.90 (Burris et al.\ 2000).}
\end{deluxetable}

\clearpage

%\section*{FIGURE CAPTIONS}

\clearpage

\clearpage

\clearpage

\clearpage

\clearpage

\clearpage

\clearpage

\clearpage

\clearpage

\clearpage

\end{document}